%% file: mypaper.tex
\documentclass{pasj00}
\draft

\begin{document}
\SetRunningHead{J. Nakashima and S. Deguchi}{SiO Maser Survey of Cold IRAS Sources}
\Received{2002/08/22}%{yyyy/mm/dd}
\Accepted{2002/11/16}%{yyyy/mm/dd}

\title{SiO Maser Survey of Cold IRAS Sources}

%%% begin:list of authors
\author{Jun-ichi \textsc{Nakashima}%
  \thanks{Present address: Department of Astronomy, University of Illinois at Urbana-Champaign, 1002 West Green Street, MC-221, Urbana, IL 61801}
  }
\affil{Department of Astronomical Science, The Graduate University for Advanced Studies,
\\ Nobeyama Radio Observatory, Minamimaki, Minamisaku, Nagano 384-1305}
\email{junichi@astro.uiuc.edu}
\and
\author{Shuji \textsc{Deguchi}}
\affil{Nobeyama Radio Observatory, National Astronomical Observatory,
\\ Minamimaki, Minamisaku, Nagano 384-1305}\email{deguchi@nro.nao.ac.jp}
%\and
%\author{C-Firstname {\sc C-Familyname}}
%\affil{C-Address of Institute}\email{ccccc@xxx.xxx.xx.xx}
%%% end:list of authors

%%% Please use the following style in case that sorting by 
%%% affilation is impossible. 
%
% \author{%
%   D-Firstname \textsc{D-Familyname}\altaffilmark{1}
%   E-Firstname \textsc{E-Familyname}\altaffilmark{1,2}
%   and
%   F-Firstname \textsc{F-Familyname}\altaffilmark{2}}
% \altaffiltext{1}{Address of Institute}
% \email{ddddd@xxx.xxx.xx.xx}
% \email{eeeee@xxx.xxx.xx.xx}
% \altaffiltext{2}{Address of Institute}

%% `\KeyWords{}' always has to be placed before `\maketitle'.
\KeyWords{Stars: late-type --- masers --- Stars: AGB and post-AGB --- Galaxy: kinematics and dynamics} 
%Do NOT move this preamble from here!

\maketitle

%%%%%%%%%%%%%%%%%%%%%%%%%
%%%%%%% ABSTRACT  %%%%%%%
%%%%%%%%%%%%%%%%%%%%%%%%%
\begin{abstract}
We present the results of observations of cold IRAS sources in the Galactic disk area, $-10^{\circ} < l < 100^{\circ}$ and $|b|<5^{\circ}$, in the SiO $J=1$--$0$, $v=1$ and 2 maser lines. SiO masers were detected in 51 out of 143 observed sources; 45 were new detections in SiO masers. The selected IRAS sources were objects with dust temperatures of between 160 and 280 K. According to a confirmation using 2MASS near-infrared images, a majority of the sample are AGB or post-AGB stars, although dense cores in the star-forming regions (or dusty H{\sc ii} regions) are involved in part of the sample. Among new detections, two were candidates for post-AGB stars: IRAS 18450$-$0148 (W 43A), and 19312$+$1950. We found that the intensity ratios of the SiO $J=1$--$0$, $v=2$ to the $v=1$ line of the objects clearly correlate with those IRAS colors. The detection rates of SiO masers tend to increase toward the Galactic center as well as the cases of previous SiO maser surveys of typical AGB stars. No strong associations of the objects to the spiral arms were found. The radial-velocity dispersion of the present sample is comparable with the dispersion of the SiO maser sample of typical AGB stars. These facts suggest that the present sample of cold IRAS sources with SiO masers has a kinematic property very similar with that of typical AGB stars. 
\end{abstract}

%%%%%%%%%%%%%%%%%%%%%%%%%%%%%
%%%%%%% INTRODUCTION  %%%%%%%
%%%%%%%%%%%%%%%%%%%%%%%%%%%%%
\section{Introduction}
 A sample of cold IRAS point sources mainly consists of evolved stellar objects with a thick dusty envelope (\cite{dav93}; \cite{lik87}; \cite{lik89}; \cite{lik91}; \cite{sil93}; \cite{tel91a}a), young stellar objects in star-forming regions, and dusty H{\sc ii} regions accompanying hot stars (\cite{pas94}; Wilking et al. 1989, 1990; \cite{yan02}). Because both late-type stellar objects and young stellar objects exhibit similar IRAS colors, it is difficult to distinguish the evolved stars from young stellar objects solely from the IRAS color indices (e.g., van der Veen et al. 1989). In addition, because considerable percentages of the IRAS 60 $\mu$m flux densities of the bulge AGB stars near the galactic plane were found to be contaminated by dust emission from surrounding interstellar clouds (\cite{izu99}; Deguchi et al. 2000a, b), the color index, $C_{23}$ (a logarithmic ratio of IRAS 60 to 25 $\mu$m flux density), may not effectively used to identify the late-type stellar objects (AGB or post-AGB stars) among these cold objects. Mass-losing O-rich AGB stars very often exhibit maser emission; OH (1612 MHz) and H$_{2}$O maser lines have been used to identify evolved stellar objects among cold IRAS sources (\cite{dav93}; \cite{lik89}; \cite{sil93}; \cite{tel91a}a). SiO masers could also be an identifier of late-type stellar objects. However, it has been found in the past SiO maser surveys that the detection rate of SiO masers are maximized at the IRAS color, $C_{12} \simeq 0$ ($T_{\textrm{dust}} \simeq 300$ K; $C_{12}$, a logarithmic ratio of IRAS 25 to 12 $\mu$m flux densities) and the rate drops quickly in $C_{12}>0.2$ ($C_{12}=0.2$ corresponds to $T_{\textrm{dust}} \simeq 280$K; e.g., \cite{jia95}; \cite{nym98}). Therefore, for IRAS objects with $C_{12} > 0.3$, SiO maser surveys have never been rigorously conducted at Nobeyama.

In the IRAS two-color diagram (e.g., \cite{van88}), typical mass-losing AGB stars occupy the region $C_{12}<0.4$ and $C_{23}<-0.1$  (region III in figure 2 in \cite{van89}). More evolved objects with thicker/hollow dust shells fall in the region  $C_{12}>0.4$ or $C_{23}>-0.1$ (regions IV and VIII). The boundary between the AGB and the post-AGB phase (proto-planetary nebula phase) is considered to lie around $C_{12} \simeq 0.5$ (e.g., \cite{nym98}; \cite{van88}). Therefore, the cold IRAS objects with SiO masers ($C_{12} \gtrsim 0.5$) are stars with a very thick dust shell, which are at a very late stage of the AGB phase (or possibly at a very early post-AGB phase). 

Proto-planetary nebulae are transient objects between the AGB phase and planetary nebulae. These objects often have colors beyond $C_{12} \sim 0.5$. Detections of SiO masers from such objects must indicate that they are at a very early phase of proto-planetary nebulae, objects of decreasing mass-loss rate, developing a high-velocity flow, and creating a hollow dust shell (\cite{eng02}). Such objects occasionally appear as large extended nebulae (the apparent size is larger than $\sim$ 10$''$) in the near-infrared, as, for example, OH 231.8+1.4 (\cite{mor87}) and IRAS 19312+1950 (\cite{nakdeg00}). However, not all of the objects with colors beyond $C_{12} \sim 0.5$ exhibit such large extended nebulosity. It is rather relatively rare to find such well-developed nebulous objects like as OH231.8+1.4 in near- and mid-infrared imaging surveys (e.g., \cite{mei99}). 

An alternative idea (\cite{suh97}) to explain the SiO masers in cold IRAS sources is that these objects with high $C_{12}$ are still at the AGB phase, and are at periods soon after a He-shell flash in a super-wind phase when a central star temporarily decreases or terminates the mass-loss  (e.g., \cite{vas93}; \cite{blo95}). Because of the highly reddened colors ($C_{12} \gtrsim 0.5$), these objects must be at a phase nearly leaving AGB (but still in O-rich; e.g., \cite{vanl01}). A temporal termination of mass-loss creates a hollow shell which exhibits a high $C_{12}$ color. A gradual recovery of mass-loss after the He-shell flash restarts SiO masers, which are emitted near the central star within a few stellar radii. These objects may mimic the colors of post-AGB stars, but may not have extended nebulosity. 

In order to find more SiO masers in cold IRAS sources and to investigate their properties, we surveyed cold IRAS objects with $T_{\textrm{dust}} = 160$--$280$ K near the galactic plane in 43 GHz SiO maser lines. The outline of the paper is as follows. In section 2, we describe the source selection, the observation, the data reduction and the results of observations. In section 3, we discuss the color characteristics of SiO detections, the intensity ratio of the $J=1$--$0$, $v=2$ to $v=1$ line and the longitude distribution of the detections. In section 4, we discuss remarkable individual objects in the sample, and finally conclude the paper in section 5.

%%%%%%%%%%%%%%%%%%%%%%%%%%%%%%%%%%%%%%%%
%%%%%%% OBSERVING SOURCES AND OBSERVATIONS %%%%%%%%
%%%%%%%%%%%%%%%%%%%%%%%%%%%%%%%%%%%%%%%%
\section{Observations}
\subsection{Source Selection}
The present sample was selected from the IRAS Point Source Catalog (version 2) in the area of the Galactic plane, $-10^{\circ} < l < 100^{\circ}$, and $|b| < 5^{\circ}$ (with one exception, IRAS 23572+6702, see table 1). The selected color ranges were $0.2 < C_{12}\equiv \log (F_{25}/F_{12}) < 0.8$, and $-1.0 < C_{23}\equiv \log (F_{60}/F_{25}) <1.5$, where $F_{12}$, $F_{25}$, and $F_{60}$ are IRAS 12, 25, and 60 $\mu$m flux densities, respectively. These color criteria effectively extract dust-enshrouded objects with $T_{\textrm{duxt}} \simeq 160$--$280$ K. The total number of observed sources was 143. Figure 1 shows the distribution of the observed sources in the Galactic coordinates. The IRAS 12 $\mu$m flux densities of observed sources range from about 2 Jy to over 1000 Jy, and most of the observed sources are brighter than 10 Jy at 12 $\mu$m.

Near-infrared imaging observations of IRAS / SiO sources at the inner disk (\cite{deg01a}a) confirmed that a considerable number of near-infrared counterparts are located at positions with large angular separations  (more than 10$''$) from the IRAS positions. Because of the high source density in the Galactic disk and bulge, the IRAS satellite might not determine the positions well, and the IRAS positions may not be sufficient to place faint mid-infrared sources in a telescope beam. Therefore, we partly used the positions in the MSX catalog instead of those in the IRAS catalog (MSX: Midcourse Space Experiment; \cite{pri97}). Since the typical astrometric uncertainty of positions in the MSX catalog is a few arcsec, it is sufficient to place mid-infrared point sources in the telescope beam of HPBW$=40''$ for the present SiO observations. We crosschecked the source positions between the IRAS and MSX catalogs, and adopted the positions of the MSX sources nearest to the IRAS positions within 40$''$ as a counterpart. The adopted MSX counterparts and the angular separation from the IRAS positions are given in table 1. We found that only three MSX counterparts have large separations (more than 20$''$) from the IRAS positions. Figure 2 shows a comparison of the flux densities between the corresponding IRAS and MSX sources, and figure 3 is the same plot for mid-infrared colors. Because the MSX fluxes and colors correlate well with those of IRAS, most of the identifications are considered to be made correctly. However, because most of the sources in the present sample are Mira-type pulsating stars (e.g., \cite{nak00}), the sample in figures 2 and 3 has a moderate dispersion, as expected from time variations of $\sim 0.5$ mag at mid-infrared wavelengths (e.g., \cite{ona97}). In fact, because the MSX catalog became available around May 2000 when we had finished the survey of 90 sources at $l>40^{\circ}$, we used the MSX positions for the survey of 53 sources at $-10^{\circ}<l<40^{\circ}$. Therefore, we list in table 1 the MSX counterparts of IRAS sources only for the sources at $-10^{\circ}<l<40^{\circ}$.

\subsection{Observations and Results}
Simultaneous observations in the SiO $J=1$--$0$, $v=1$ and $2$ transitions at 43.122 and 42.821 GHz, respectively, were made with the 45-m radio telescope at Nobeyama during the periods of 2000 April--May and 2001 March. In addition to the SiO lines, observations in the H$_2$O, $6(1,6)$--$5(2,3)$ transition at 22.235 GHz were made for the SiO maser emitters, which were found during the present SiO observations. The beam size of the telescope was about 40$''$ at 43 GHz and 72$''$ at 22 GHz. In the SiO observations, a cooled SIS receiver with a bandwidth of about 0.4 GHz was used, and the system noise temperatures (including atmospheric noise and antenna ohmic loss) were about 200--300 K (SSB), depending on the weather conditions. In the H$_2$O observations, a cooled HEMT receiver with a bandwidth of about 2 GHz was used, and the system noise temperatures were about 170--300 K. The aperture efficiency of the telescope was about 0.57 at 43 GHz and 0.62 at 22 GHz. The conversion factor from the antenna temperature to the flux density was 2.9 Jy K$^{-1}$ at 43 GHz and 3.0 Jy K$^{-1}$ at 22 GHz. Acousto-optical spectrometer arrays of high and low resolutions (AOS-H and AOS-W) were used. The AOS-H spectrometer has a 40 MHz bandwidth and 2048 frequency channels, with an effective spectral resolution of 0.29 km s$^{-1}$ at 43 GHz and 0.50 km s$^{-1}$ at 22 GHz. Likewise, the AOS-W spectrometer has a 250MHz bandwidth and 2048 frequency channels, with an effective spectral resolution of 1.7 km s$^{-1}$ at 43 GHz and 3.4 km s$^{-1}$ at 22 GHz. We used multiple AOS-H spectrometers to cover a wide velocity range at a high frequency resolution. Because the line widths of masers are often less than 1 km s$^{-1}$, we need the high resolution achieved by the AOS-H. The radial-velocity coverage was $\pm$ 350 km s$^{-1}$ at 43 GHz and $\pm$600 km s$^{-1}$ at 22 GHz. This entire velocity range was covered by the AOS-H spectrometers (AOS-W was used to confirm the detections). All of the observations were made in position-switching mode using 10$'$ off-positions at both 43 and 22 GHz. The pointing accuracy of the telescope was checked every 2 or 3 hours by 5-point mapping of nearby SiO maser sources, V1111 Oph and $\chi$ Cyg. The pointing accuracy was found to be better than 5$''$.

Raw data were processed by flagging out bad scans, making r.m.s.-weighted integrations, and removing slopes in the baselines. Detections of maser lines were judged by the following criteria. For narrow spike-type emissions, the peak antenna temperature must be greater than the 3 $\sigma$ level of the r.m.s. noise. (For extremely narrow lines with a bandwidth less than 1 km s$^{-1}$, we carefully checked individual scans in raw data to exclude spurious lines.) For broad emissions, the effective signal-to-noise ratio (S/N) over the line width must be larger than 10; the effective S/N was calculated from integrated intensities and the r.m.s. noise. Furthermore, we inspected each spectrum by eye and discarded some marginal detection that satisfied the above criteria. 

Finally, we obtained 51 detections out of 143 observed sources in either SiO $J=1$--$0$, $v=1$ or 2 transitions; 45 of 51 detected sources were new detections in SiO masers, and 9 of 45 new detections had never been detected in OH or H$_{2}$O maser lines. The results of our observations of SiO masers are summarized in table 2 for detections and in table 3 for non-detections. The spectra of the detections are shown in figure 4. The velocities, $V_{\textrm{lsr}}$, in table 2 are the radial velocities at the intensity peaks of spectra of SiO masers. In the case of the H$_2$O maser line, we detected 3 out of 23 observed sources; only one of them (IRAS 19312+1950) was a new detection in the H$_2$O maser line (e.g., \cite{nakdeg00}). The H$_2$O maser spectra are shown in figure 5. All of the detected H$_2$O masers were recognized as "Type B" profiles (double peak; e.g., \cite{eng86}). At first glance, the H$_2$O maser of IRAS 19312+1950 (see figure 1 in \cite{nakdeg00}) seems to be like "Type A" (single peak at the stellar velocity), but should be regarded as "Type B", because we confirmed the other velocity component at $V_{\textrm{lsr}} \sim 60$ km s$^{-1}$ by a recent observation (Deguchi and Nakashima, in preparation). The line parameters are summarized in table 4 for H$_2$O detections. The results of non-detections are listed in table 5.

%%%%%%%%%%%%%%%%%%%%%%%%%%
%%%%%%% DISCUSSION %%%%%%%
%%%%%%%%%%%%%%%%%%%%%%%%%%
\section{Statistical Properties and Discussions}
\subsection{Color Characteristics of SiO Detections}
Figure 6 shows a two-color diagram of the observed sources. The broken lines in figure 6 indicate the boundaries of the source types suggested by \citet{van88}. The observed sources distribute mostly in regions IIIb, IV, and VIII, which are mainly composed of variable AGB stars with thick O-rich circumstellar shells, variable AGB stars with very thick O-rich circumstellar shells, and other kinds of very cool objects, respectively (\cite{van88}). SiO detections clearly concentrate into the regions IIIb and IV. This phenomenon supports previous findings that SiO masers are emitted mostly from AGB stars with an O-rich dust envelope (e.g., \cite{hab96}). On the other hand, a few SiO detections have been confirmed in regions VIII and V (mostly in region VIII). 

Region VIII involves AGB stars, post-AGB stars, young stellar objects (dense cores in star-forming regions), and galaxies (\cite{van88}; \cite{stu91}; \cite{pas94}). About 60\% of the LRS spectra of bright IRAS sources in region VIII show weak-to-moderate 10 $\mu$m absorption. Then, about 30\% of the LRS spectra are typical for planetary nebulae (\cite{van88}). The associations of IRAS sources with previously known objects were also studied by \citet{van88}; 3\% of the IRAS sources in region VIII are associated with a planetary nebula and 4\% with an early-type star with emission lines. A relatively large fraction (15\%) is associated with galaxies. However, most IRAS galaxies exhibit 12 $\mu$m flux densities of less than 2 Jy (\cite{ful89}), which equals the lower limit of the 12 $\mu$m flux density of the present sample. Therefore, the possibility of the galaxies in the sample is excluded. Because young stars (or dense cores in star-forming regions) concentrate in the region of color larger than $C_{12}\simeq 1.0$ in the two color diagram (\cite{stu91}; \cite{pas94}), a majority of the objects in the sample would be late-type stars (AGB or post-AGB stars). 

In the present sample, it would be appropriate that SiO-detected sources are classified as late-type stars (AGB or possibly post-AGB stars). SiO masers in star-forming regions are quite rare (\cite{bar84}; \cite{jew84}). In fact, more than 99\% of the SiO maser emitters previously known originate from late-type stars, except 3 extreme objects in star-forming regions: Orion IRc 2, W 51 IRs 2, and Sgr B2 MD5 (\cite{has86}; \cite{fue89}; \cite{mori92}). These objects in star-forming regions with SiO masers are surrounded by largely extended mid-infrared nebulosity. To check the contamination of young stellar objects in SiO detections, we inspected mid-infrared images toward SiO detections using MSX image archives with about 0.3' resolution (\cite{pri97}), but no extended source has been found toward SiO detections. In addition to the MSX images, we inspected near-infrared $J$, $H$, $K$-bands images toward observed sources using the 2MASS image archive. The results of the inspection are given in table 1. Most objects with SiO masers were point-like. Some SiO detections were too faint to be seen in 2MASS near-infrared images. We found two objects with a large extended nebulosity: IRAS 19312+1950 and 20249+3953. Although the morphology of IRAS 20249+3953 in the 2MASS images was point symmetric (typically seen in post-AGB stars), we finally concluded that this object is a dense core in a star-forming region. An extended source and a dark lane were seen toward IRAS 20249+3953 in MSX and DSS images, respectively. About 10\% of the present sample was recognized as star-forming regions using 2MASS images, and were non-detections in SiO masers.

The color index, $C_{12}$, which is believed to be an indicator of the evolutionary stage of AGB stars, increases with the mass-loss rate. Near the end of the AGB phase, the mass-loss rate often increases enormously (up to 10$^{-4}$ $M_{\odot}$ y$^{-1}$; \cite{blo95}; \cite{yam96}). The expansion velocity also increases ($\sim$ 20 km s$^{-1}$); it is called superwind (\cite{ren81}). The termination of a superwind is considered to be the end of the AGB phase (\cite{ren81}). The boundary between the AGB and post-AGB phases is considered to be roughly at $C_{12} \sim 0.5$ in the two-color diagram (e.g., Figure 3 in \cite{nym98}; \cite{vanh97}). After the mass-loss (and pulsation) ceases, circumstellar gases are not supplied from the central star; SiO masers disappear first among the circumstellar masers (OH, H$_{2}$O and SiO) in the O-rich circumstellar envelope, because SiO masers are emitted within a few stellar radii from the central star. Therefore, when a star goes into the post-AGB phase, SiO masers would abruptly blow out to cease. In other words, if the IRAS color is an evolutionary indicator, a clear edge of the distribution of SiO emitters would be recognized in an IRAS two-color diagram, which corresponds with the boundary between AGB and the post-AGB phase. In fact, the edge of the SiO maser distribution can be seen around $C_{12} \sim 0.5$ in figure 6, though it is obscured by contamination of young stellar objects. 

According to above perspective, SiO emitters with IRAS colors just beyond $C_{12}\sim 0.5$ can be regarded as stars at an early post-AGB phase (or at a possibly very late AGB phase) decreasing/ceasing mass-loss. We found three SiO emitters with the IRAS colors, $C_{12} > 0.5$, in the present sample: IRAS 18135$-$1456 (OH15.7+0.8), 18450$-$0148 (W 43A), and 19312+1950. Two of them (IRAS 18450$-$0148 and 19312+1950) were the first detections in SiO masers. An SiO maser toward IRAS 18135$-$1456 was previously detected by \citet{nym93}. We detected SiO masers toward IRAS 18498$-$0017, which is also considered to be a candidate for a post-AGB star because of its color, $C_{12}=0.46$. We suggest that SiO masers of these stars (except IRAS 18498$-$0017) are still alive in the "dying AGB wind" during the mass-loss reduction process accompanying the departure from the AGB phase (e.g., \cite{eng02}). The IRAS variability indices for IRAS 18135$-$1456 and 18450$-$0148 are 0 and 19, respectively, indicating that these two are not at the mira-type pulsating phase (almost equal to AGB phase), but at the post-AGB phase. The H$_{2}$O maser of IRAS 18135$-$1456 has tended to gradually fade during last 10 years; after 1993, the H$_{2}$O maser was detected only once in 1997 (\cite{eng02}). The velocity width of the H$_2$O maser emission of IRAS 18450$-$0148 ($\sim$ 170 km s$^{-1}$, \cite{lik92}) exceeds that of the OH emission ($\sim$ 14 km s$^{-1}$, \cite{tel89}), indicating that the objects have recently become a post-AGB star (\cite{gom94}). In the cases of IRAS 18498$-$0017 and 19312+1950, the IRAS variability indices are 96 and 77, respectively. The index for IRAS 18498$-$0017 is somewhat too high to be regarded as a post-AGB star. Although H$_2$O and OH masers have been detected toward this object, the profiles of masers are like normal double peaks typically seen in AGB stars (\cite{tel89}; \cite{eng86}). A relatively high value of the IRAS variability index of IRAS 19312+1950 might mean that this object is still at the AGB phase. However, \citet{nakdeg00} reported that IRAS 19312+1950 has an extended near-infrared counterpart, which is typically seen in proto-planetary nebulae.

In order to check the contamination by interstellar dust in the IRAS 60 $\mu$m flux densities, we made color--latitude diagrams, plots of IRAS colors, $C_{12}$ and $C_{23}$, against the galactic latitudes, as shown in figure 7. In both panels in the figure 7, clear concentrations of the sources toward the galactic plane ($b=0^{\circ}$) can be seen. This concentration would be partly due to contamination by young stellar objects which are bound within $|b|\leq 1^{\circ}$. Another remarkable characteristic in figure 7 is a clear separation in the distribution between SiO detections and non-detections. IRAS colors of SiO detections do not strongly depend on the galactic latitudes except a few post-AGB candidates; this means that the IRAS colors well reflect the colors of the objects themselves in the case of SiO detections. In the upper panel of figure 7, the concentration of non-detections (open circles) around $C_{23}\sim 1$ is remarkable. Some of these sources are, in fact, known dense cores in star-forming regions. According to 2MASS near-infrared images, approximately 10\% of the present sample is inferred to be young stellar objects in star-forming regions. Generally speaking, star-forming regions are recognized as a dense star cluster or an extended H{\sc ii} region in 2MASS images. In the present research, no SiO maser was detected toward young stellar objects.

The bulge SiO maser sources distribute quite differently in the color-latitude diagram; the color indices, $C_{23}$, of the SiO detections increase steeply up to $\sim 0.5$ near $|b|\sim 0^{\circ}$ (see figures 4 in Deguchi et al. 2000a, b). The SiO detections in the upper panel of figure 7 do not show such a behavior, except candidates for post-AGB stars with SiO masers (indicated by arrows in figure 7). Of course, post-AGB stars are supposed to have colors with higher $C_{12}$ and $C_{23}$ than do the other SiO detections. Therefore, we conclude that the IRAS colors, $C_{23}$ of SiO detections in the present sample, are intrinsic to the sources, and that the contamination by interstellar dust in the IRAS 60 $\mu$m flux densities in the present sample is relatively small. This is probably because the IRAS sources in the present sample are relatively bright objects ($F_{12} \geq 10$ Jy) compared with those in the previous bulge samples (Deguchi et al. 2000a, b).

\subsection{Intensity Ratio of the $J=1$--$0$, $v=2$ to $v=1$ Lines}
The intensity ratio of the SiO $J=1$--$0$, $v=2$ to the $J=1$--$0$, $v=1$ lines of AGB stars has been observed as being nearly equal to unity (\cite{sch79}; \cite{spe81}; \cite{naka93}). Likewise, past results of SiO maser observations toward IRAS sources, which are normal AGB stars, have indicated that the intensity ratio was roughly unity (e.g., \cite{izu94}). At a glance, it seems to be difficult to explain the intensity ratio of near unity of the two rotational lines in the different vibrational states with very different excitation temperatures as $T_{\textrm{ex}}=1600$ K ($v=1$) and $T_{\textrm{ex}}=3200$ K ($v=2$; e.g., \cite{eli92}). In order to explain it, Olofsson et al. (1981, 1985) proposed a line overlapping between the infrared transition of SiO (from $v=1$, $J=0$ to $v=2$, $J=1$) and a line of the H$_2$O molecule ($\nu_{2}$=0 12$_{75}$) -- ($\nu_{2}$=1 11$_{66}$), making a pump cycle between the SiO energy levels; $J=$0 $v=$1 $\Rightarrow$ $J=$1 $v=$2 $\rightarrow$  $J=$0 $v=$2 $\Rightarrow$ $J=$1 $v=$1 $\rightarrow$  $J=$0 $v=$1, where vibration--rotation and maser transitions are indicated by $\Rightarrow$ and $\rightarrow$, respectively. This pump cycle, if it works, leads to a relative strengthening of the $v=2$ $J=1$--$0$ maser in comparison with the case of the absence of a pump cycle (\cite{olo85}). On the other hand, however, there are many exceptional cases. For example, that the intensity ratio of IRC$-$10414 is less than about 0.1 (\cite{ima99}).  \citet{nym86} suggested that the average intensity ratio of SiO masers from OH/IR stars were nearly 4. 

In the present research, we found a clue linking these intensity ratios of SiO masers. Figure 8 shows a plot of intensity ratios of the $J=1$--$0$, $v=2$ to the $J=1$--$0$, $v=1$ line (using the integrated and peak intensities) against the IRAS color index, $C_{12}$. The filled marks show the data in the present observation and the open marks indicate the data from a previous SiO maser survey of the inner-disk IRAS sources with $C_{12}<0.2$ (\cite{nakdeg02}). Both of the sources are located in almost the same galactic longitude range. The triangles and inverse triangles show the lower and upper limits, respectively. We can see that the $v=$2/1 line-intensity ratio tends to increase with the IRAS color of the objects until $C_{12}=0.5$. The average values of the intensity ratios of the open circles in figure 8 are nearly unity (the logarithmic values are nearly 0). In contrast, in cold IRAS sources with SiO masers (filled circles, $0.2<C_{12} < 0.5$), the $J=1$--$0$ $v=$2 line, tends to be much stronger than the $J=1$--$0$ $v=$1 line as shown in figure 8. Beyond $C_{12} \sim 0.5$, the intensity ratios tend to decrease. The least-square best fits in the color range of $0.2<C_{12} < 0.5$ (without upper and lower limits) in figure 8 give
%%%equation 1%%%
\begin{equation}
   \log (F_{v=2}/F_{v=1}) = 0.88\,(\pm 0.13)\, C_{12} + 0.06\, (\pm 0.03)
\end{equation}
%%%%%%%%%%%%%%\\
for the integrated intensity ratio, and
%%%equation 2%%%
\begin{equation}
   \log (T_{v=2}/T_{v=1}) = 0.56\,(\pm 0.07)\, C_{12} + 0.07\, (\pm 0.01)
\end{equation}
%%%%%%%%%%%%%%\\
for the peak-intensity ratio. Simple correlation coefficients are about 0.5 for both of the upper and lower panels in figure 8.

In the present observations, both SiO transitions, $J=1$--$0$, $v=1$ and 2, are near to the band edges of the SIS receiver with a band width of about 0.4 GHz. Therefore, the sensitivity at each frequency varies depending on the daily tuning condition of the receiver. Therefore, we carefully checked the reliability of the intensity calibration at different nearby frequencies using pointing sources and a molecular cloud, Sgr B2 every day, and confirmed that the intensity calibrations of all spectrometers were nearly equal through all of the observing sessions. \citet{nym93} made a same plot with figure 8 using the ($K$--$L$) color as a horizontal axis, and found same correlation between the intensity ratio and the ($K$--$L$) color. The discrepancy of the correlation (in our case, seen at $C_{12}=0.5$) could not be seen in Nyman et al.'s(1993) result.

Here, we would like to suggest an interesting resemblance with SiO masers in star-forming regions: Orion IRc 2, W 51 IRs 2, and Sgr B2 MD5 (\cite{has86}; \cite{fue89}; \cite{mori92}). The line intensity ratios in the latter two SiO masers in star-forming regions greatly deviate from unity. The intensity of the $J=1$--$0$, $v=1$ line from Sgr B2 is about 1.5 Jy, while that of the $J=1$--$0$, $v=2$ line is less than 0.3 Jy (\cite{shi97}). In the case of W 51 IRs2, the $J=1$--$0$, $v=2$ intensity is more than 10-times stronger than that of $J=1$--$0$, $v=1$. Among these three objects, only Sgr B2 MD5 is included in the IRAS point source catalog (IRAS 17441$-$2822) and the IRAS colors are available [($C_{12}$, $C_{23}$)=(0.95, 2.28)]. These colors of Sgr B2 MD5 might suffer from serious contamination by surrounding material. According to mid-infrared photometry for the Ori KL region with SiO masers by \citet{gez98}, the mid-infrared color of IRc2 (a logarithmic ratio of the 12 to 20 $\mu$m intensity) is about 0.2. For the case of W 51 IRs2, it is somewhat difficult to estimate the mid-infrared fluxes because this source consists of at least 7 continuum sources; four of them are ultra-compact H{\sc ii} regions (\cite{oka01}). For the moment, we estimate the color of W 51 IRc2 to be $C_{12}\sim$ 0.6 by extrapolating the spectrum in figure 2 of \citet{oka01} to longer wavelengths. Therefore, the increasing color sequence of these three sources is "Orion IRc 2---W 51 IRs 2---Sgr B2 MD5". The observed line-intensity ratios to the mid-infrared color for these three sources are consistent with the correlation shown in figure 7. The ratio increases with $C_{12}$, but decrease beyond $C_{12}\sim 0.5$. Therefore, the line intensity-color correlation found in the evolved stellar objects also seems to be applied to the SiO maser sources in the star-forming regions, and may be useful to deduce the pumping mechanism of SiO masers. 

\subsection{Longitude Distribution}
Figure 9 shows a histogram of the galactic longitudes for the SiO detections (shadow) and non-detections (blank). The detection rates are also shown by solid line with filled circles. For a comparison, the detection rates in the previous SiO maser surveys of sources with IRAS colors of typical AGB stars are also shown by the dotted line with filled squares (\cite{izu99}; \cite{deg00a}a; Deguchi et al. 2000a, b; \cite{nakdeg02}; \cite{nak02}). The detection rate in this sample tends to decrease with the galactic longitude, as found in previous SiO surveys of IRAS sources with the color $0.0 < C_{12} < 0.1$. We then see a sudden increase in the detection rate in the range between $l=20^{\circ}$ and $40^{\circ}$.  Because we used the MSX positions instead of the IRAS positions in the inner region ($l<40^{\circ}$), the difference in the survey condition can possibly be a cause of the sudden change of the detection rates; the detection rate is 55\% for 53 sources in $l<40^{\circ}$, and 24\% for 90 sources in $l>40^{\circ}$. However, the improvement of the positions is not enough to make the overall change in the detection rates above and below $l=40^{\circ}$, because the number of MSX counterparts with a large separation ($>20''$ ) from the IRAS position was only 3 in the present sample. The maximum of the detection rate between $l=20^{\circ}$ and $40^{\circ}$ may be caused by a statistical fluctuation due to the small numbers of observed sources in the bin.

Figure 10 shows a longitude--velocity diagram for SiO detections overlaid on the CO $J=1$--$0$ line map (taken from \cite{dam01}). The filled and open circles indicate the sources with distances (luminosity distance; see next paragraph) below and above 5 kpc. The spread of the radial velocities of the SiO sources seems to be within the velocity limits produced by galactic rotation, which is indicated by the CO gas distribution [except IRAS 17545$-$2512 at ($l=4 3^{\circ}\hspace{-4.5pt}.\hspace{.5pt}56$, $V_{\textrm{lsr}}=-126$ km s$^{-1}$, and $D_{L}=4.4$ kpc)]. The upper edge of the SiO distribution corresponds to the Sagittarius--Carina arm and the Scutum--Crux arm at a distance of 4--5 kpc, and the lower edge (between $l=30^{\circ}$ and $100^{\circ}$) corresponds to the outer arm at a distance of 8--10 kpc. We found that only one source in our sample, IRAS 17192$-$3206 at $l=-5 3^{\circ}\hspace{-4.5pt}.\hspace{.5pt}4$, has a large negative velocity; this source is probably associated with the central nuclear disk which is rapidly rotating, as seen in the overlaid CO map.

Figure 11 shows the distribution of SiO detections projected onto the Galactic plane. The SiO sources in the present sample do not show any systematic patterns in figure 11. As in previous papers, the distances (luminosity distance) were estimated from the bolometric fluxes of sources which were computed from IRAS 12 and 25 $\mu$m flux densities and a bolometric correction for O-rich stars (van der Veen and Breukers 1989) based on the assumption of a constant luminosity of  $8 \times 10^{3} L_{\odot}$. For post-AGB stars, the bolometric flux tends to be underestimated in the bolometric correction, because the spectral energy distributions of post-AGB stars are weighted at shorter and longer wavelengths, which is different from those of AGB stars. Because the AGB stars have a single-peak spectral energy distribution, which is peaked at a mid-infrared wavelength, the bolometric correction is relatively accurate. In contrast, post-AGB stars have a doubly peaked SED, peaked at the mid and near-infrared wavelengths (e.g., \cite{fuj02}). To make more precise bolometric corrections for post-AGB stars, we need to obtain near-infrared fluxes. 
 
We also calculated the radial velocity dispersion (root mean squares of residues subtracted by circular and solar motions from $V_{\textrm{lsr}}$) for the 22 SiO sources in $40^{\circ} < l < 70^{\circ}$, and obtained the dispersion 27.9 km s$^{-1}$. Here, a flat rotation curve (see caption of figure 7 in \cite{nakdeg02}) and the galactocentric distance of the Sun (8.5 kpc) are assumed in the calculation. This value is comparable with the dispersion, $\sim$30--35 km s$^{-1}$, for the SiO sources with the bluer IRAS colors ($C_{12}<0.2$) in several galactocentric distance ranges (\cite{nakdeg02}). This value is also comparable with the dispersion of $\sim$30 km s$^{-1}$ for the nearby SiO maser sources in the southern hemisphere (\cite{deg01b}b). For the subsample of 29 sources in $l < 40^{\circ}$, we obtained a slightly larger radial velocity dispersion, 35 km s$^{-1}$. This subsample might give a larger velocity dispersion caused by the errors involved in the distance estimations.

%%%%%%%%%%%%%%%%%%%%%%%%%%
%%% Individual Sources %%%
%%%%%%%%%%%%%%%%%%%%%%%%%%
\section{Individual Sources}
Among the observed sources, the known young stellar objects in star-forming regions are IRAS 17271$-$3309 (\cite{bro96}), 18196$-$1331 (\cite{min91}; \cite{kas92}; \cite{geb96}), 18551$-$0146 (\cite{wat99}), 19089+1542 (\cite{the94}; \cite{mee01}), 19598+3324 (\cite{ces88}), 20187+4111 (\cite{hil95}). Notable evolved stars are as follows;

\subsection*{IRAS 17295$-$3321 (OH 354.76$-$0.06), IRAS 17317$-$3331 (V1018 Sco)}
These sources were studied as maser sources associated with a globular cluster, Liller 1. However, because of radial velocity differences (\cite{sev97}; \cite{tel91b}b; \cite{tel89}), \citet{fra94} concluded that these are not associated with Liller 1. SiO masers were detected in the present research, and radial velocities coincided with those of OH masers. 

\subsection*{IRAS 18016$-$2743}
The OH  1612 MHz maser was detected (\cite{tel91b}b; \cite{dav93}; \cite{sev97}), but SiO maser searches have been negative in the $J=2$--$1$ $v=$1 line (\cite{nym98}) and in the $J=1$--$0$ $v=$1 and 2 lines in the present paper. This object was classified as a proto-planetary nebula by \citet{nym98}.

\subsection*{IRAS 18071$-$1727 (OH 12.8 +0.9)}
The OH 1612 MHz maser was detected from this object (\cite{tel89}; \cite{dav93}). SiO maser searches ($J=1$--$0$ $v=$1 and 2, and  $J=2$--1 $v=$1) have been negative (\cite{nym98}; this paper). This source was not resolved ($\leq 1''$) in mid-infrared imaging observations (\cite{mei99}).  \citet{nym98} classified this source as a proto-planetary nebula.

\subsection*{IRAS 18450$-$0148 (W 43A)}
The velocity width of H$_2$O maser emission of this source ($\sim$ 170 km s$^{-1}$, \cite{lik92}) exceeds that of the OH maser emission ($\sim$ 14 km s$^{-1}$, \cite{tel89}), indicating that this source recently entered the proto-planetary phase (e.g., \cite{gom94}). \citet{ima02} found a collimated and precessing jet of molecular gas that is traced by H$_2$O maser spots.

\subsection*{IRAS 19069+0916}
Both the SiO and H$_2$O masers were detected in this research. SiO emission was found at $V_{\textrm{lsr}}=$30.0 km$^{-1}$, though H$_2$O masers were detected at $V_{\textrm{lsr}}=-25.7$ and 10.7 km$^{-1}$ in the present paper (figure 5). The OH 1612 MHz double peaks were found at $V_{\textrm{lsr}} = 10$ and 54 km s$^{-1}$  (\cite{ede88}), giving the stellar velocity at $V_{\textrm{lsr}}\sim$32 km s$^{-1}$; this is consistent with the SiO radial velocity. The low-velocity component at $V_{\textrm{lsr}} = -26$ km s$^{-1}$ of H$_2$O maser emission, which appears lower than the velocity expected from the OH double peaks, had not been detected in the past (\cite{eng96}). It might indicate an activity of the central star at a transient stage to the post-AGB phase.

\subsection*{IRAS 19312+1950}
A large extended nebulosity can be recognized in 2MASS images toward this source. \citet{nakdeg00} found SiO and H$_2$O maser emissions. They also reported tentative detections of HCN and SO thermal lines toward this source.

\subsection*{IRAS 19327+3024 (BD+30$^{\circ}$3639)}
This is a planetary nebula, and has been rigorously observed from X-ray to radio wavelengths, including optical imaging with the HST (\cite{har97}). The relatively low temperature of the central star ($T \sim 40000$ K) and the small expansion time scale of the nebula (900 yr; \cite{kaw96}) suggests a recent emergence to the planetary-nebula phase. No maser has ever been detected.

%%%%%%%%%%%%%%%%%%%%%%%%%%
%%%%%%% CONCLUSION %%%%%%%
%%%%%%%%%%%%%%%%%%%%%%%%%%
\section{Conclusion}
We observed 143 cold IRAS sources in the SiO $J=1$--$0$, $v=1$ and 2 lines, and detected 51 objects; 45 of 51 are new detections. SiO masers were newly detected in two candidates for post-AGB stars: IRAS 18450$-$0148 (W 43A) and 19312+1950. We also detected 3 H$_2$O maser emitters out of 23 observed sources; one of them was a new detection. The main results of this research are: 

\begin{enumerate}
\item The intensity ratio of the SiO $J=1$--$0$, $v=2$ to the $J=1$--$0$, $v=1$ lines is correlated with the IRAS color for objects with $C_{12} \lesssim 0.5$.  Beyond $C_{12} \sim 0.5$, the intensity ratio dramatically drops. 

\item The IRAS colors, $C_{12}$ and $C_{23}$, of SiO emitters near the Galactic plane do not strongly depend on the galactic latitude, though the colors of SiO non-detections tend to vary with the latitude. This fact indicates that young stellar objects contaminate the sample, and that the IRAS colors of SiO emitters reflect the intrinsic colors of the objects themselves. 

\item The SiO sources are not strongly associated with the spiral arms in the longitude--velocity diagram and on the Galactic-plane map. The velocity dispersion of the SiO detections in the sub sample ($l>40^{\circ}$) is comparable with the dispersions found in previous SiO surveys for sources with different colors (typical colors of normal AGB stars). These facts suggest that the cold IRAS sources with SiO masers has a kinematic property very similar to that of typical AGB stars with SiO masers.
\end{enumerate}
\onecolumn

%%%%%%%%%%%%%%%%%%%%%%%%%%%%%%%
%%%%%%% ACKNOWLEDGEMENT %%%%%%%
%%%%%%%%%%%%%%%%%%%%%%%%%%%%%%%
The authors thank staff of Nobeyama radio observatory for the help of observations and also thank Y. Ita and T. Soma for the help in the data reductions. This research is made use of the SIMBAD database, operated at CDS, Strasburg, France and also made use of the 2MASS and MSX data archive, operated at IRSA, NASA. This paper is a part of the JN's thesis presented for the award of Ph. D. to the Graduate University for Advanced Studies.

%%%%%%%%%%%%%%%%%%%%%%%%%%%%%%%
%%%%%%% FIGURE CAPTIONS %%%%%%%
%%%%%%%%%%%%%%%%%%%%%%%%%%%%%%%

\newpage
\begin{figure}
  \begin{center}
    \FigureFile(120mm,10mm){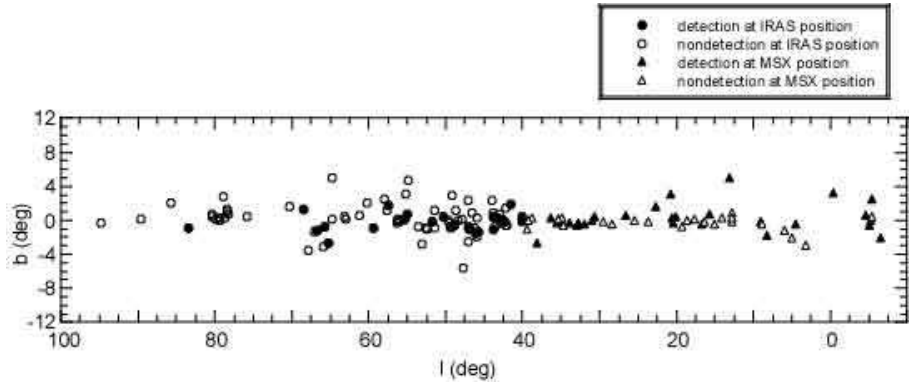}
    %%% \FigureFile(width,height){filename}
  \end{center}
  \caption{Positions of the observed sources in the galactic coordinates. The circles and triangles indicate sources observed at IRAS and MSX positions, respectively.  The filled and open marks indicate SiO detections and nondetections, respectively.}\label{fig:sample}
\end{figure}

\newpage
\begin{figure}
  \begin{center}
    \FigureFile(120mm,10mm){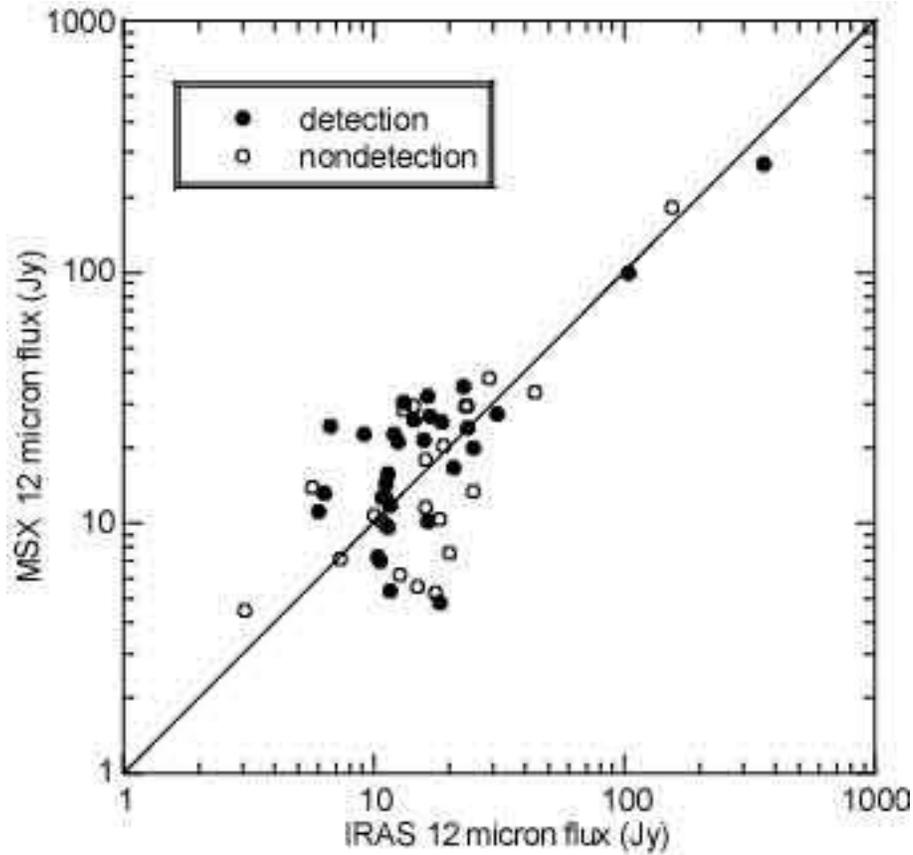}
    %%% \FigureFile(width,height){filename}
  \end{center}
  \caption{Correlation between IRAS and MSX 12 $\mu$m flux densities 
  for the sources with $l < 40^{\circ}$.}\label{fig:sample}
\end{figure}

\newpage
\begin{figure}
  \begin{center}
    \FigureFile(120mm,10mm){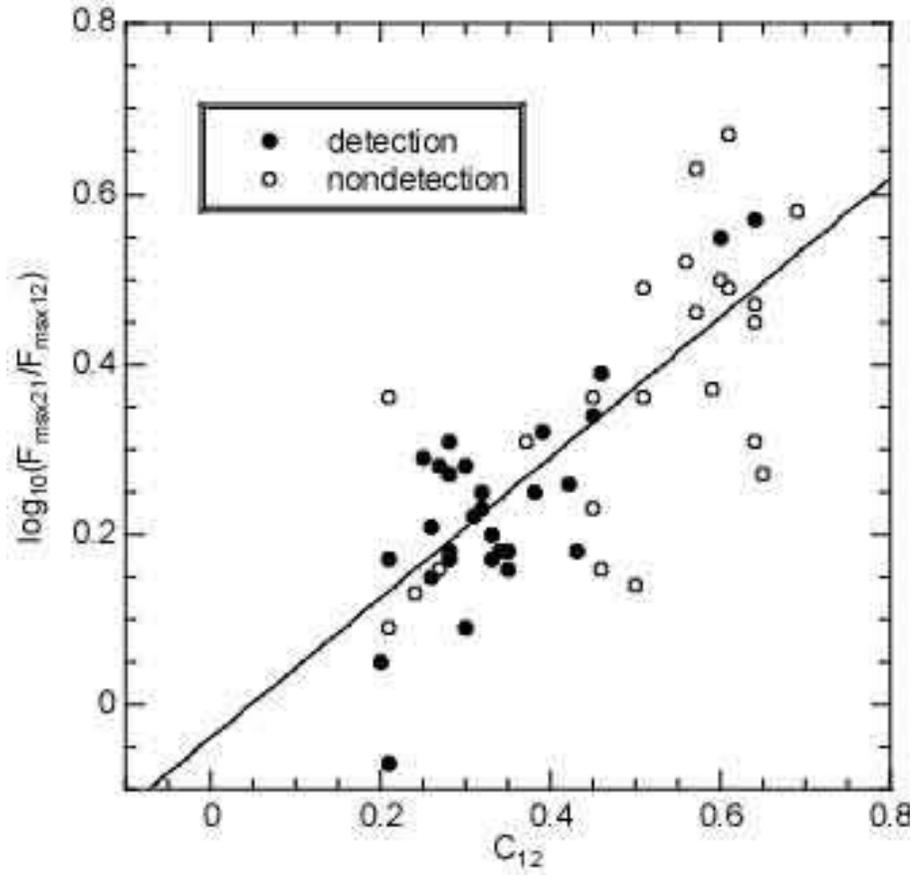}
    %%% \FigureFile(width,height){filename}
  \end{center}
  \caption{Correlation between IRAS and MSX colors for the sources with $L<40^{\circ}$. $F_{\textrm{msx12}}$ and $F_{\textrm{msx21}}$ indicate the MSX flux densities at 12 (C-band) and 21 (E-band) $\mu$m.}\label{fig:sample}
\end{figure}

\newpage
\renewcommand{\thefigure}{4a}
\begin{figure}
  \begin{center}
    \FigureFile(120mm,10mm){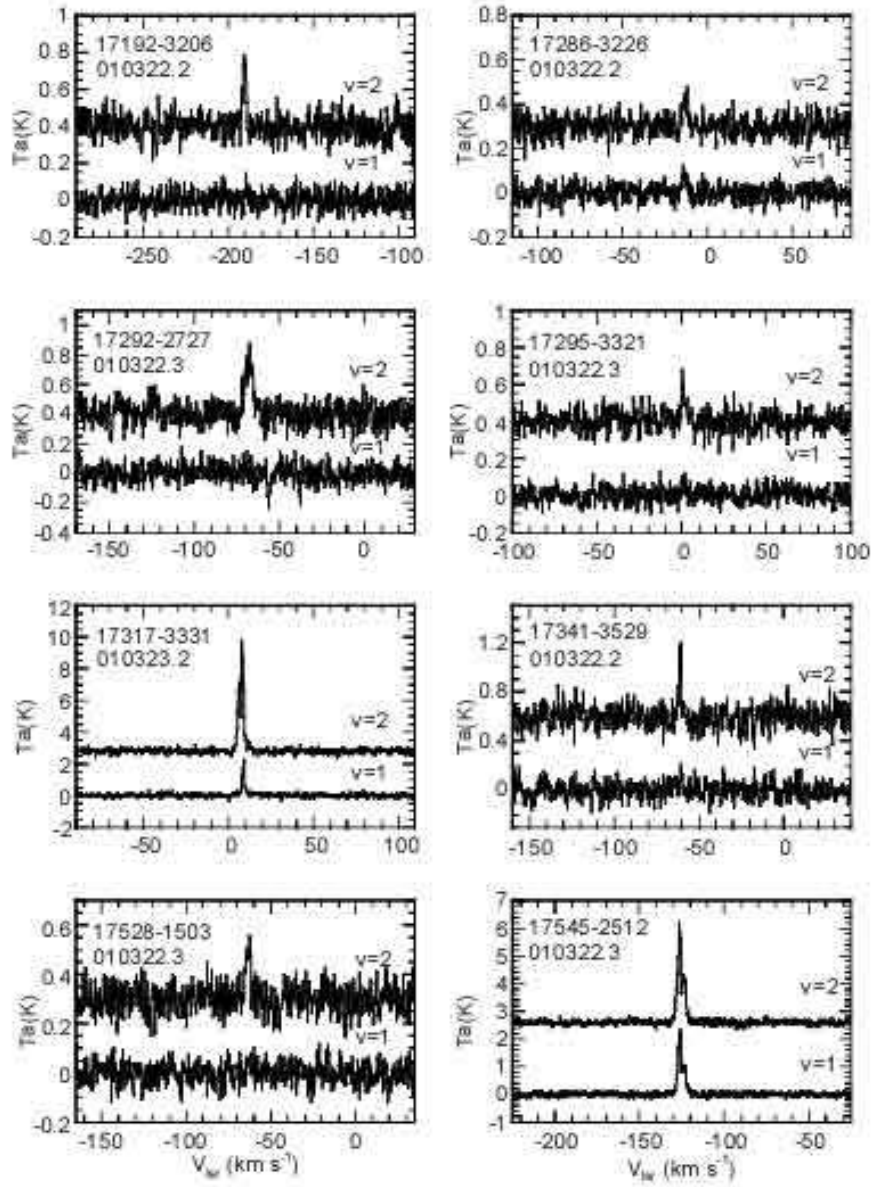}
    %%% \FigureFile(width,height){filename}
  \end{center}
  \caption{Spectra of the SiO $J=1$--$0$, $v=1$ and 2 lines for 50 of 51 detected sources, except for IRAS 19312+1950 (e.g., \cite{nakdeg00}) }
  \label{fig:sample}
\end{figure}

\newpage
\renewcommand{\thefigure}{4b}
\begin{figure}
  \begin{center}
    \FigureFile(120mm,10mm){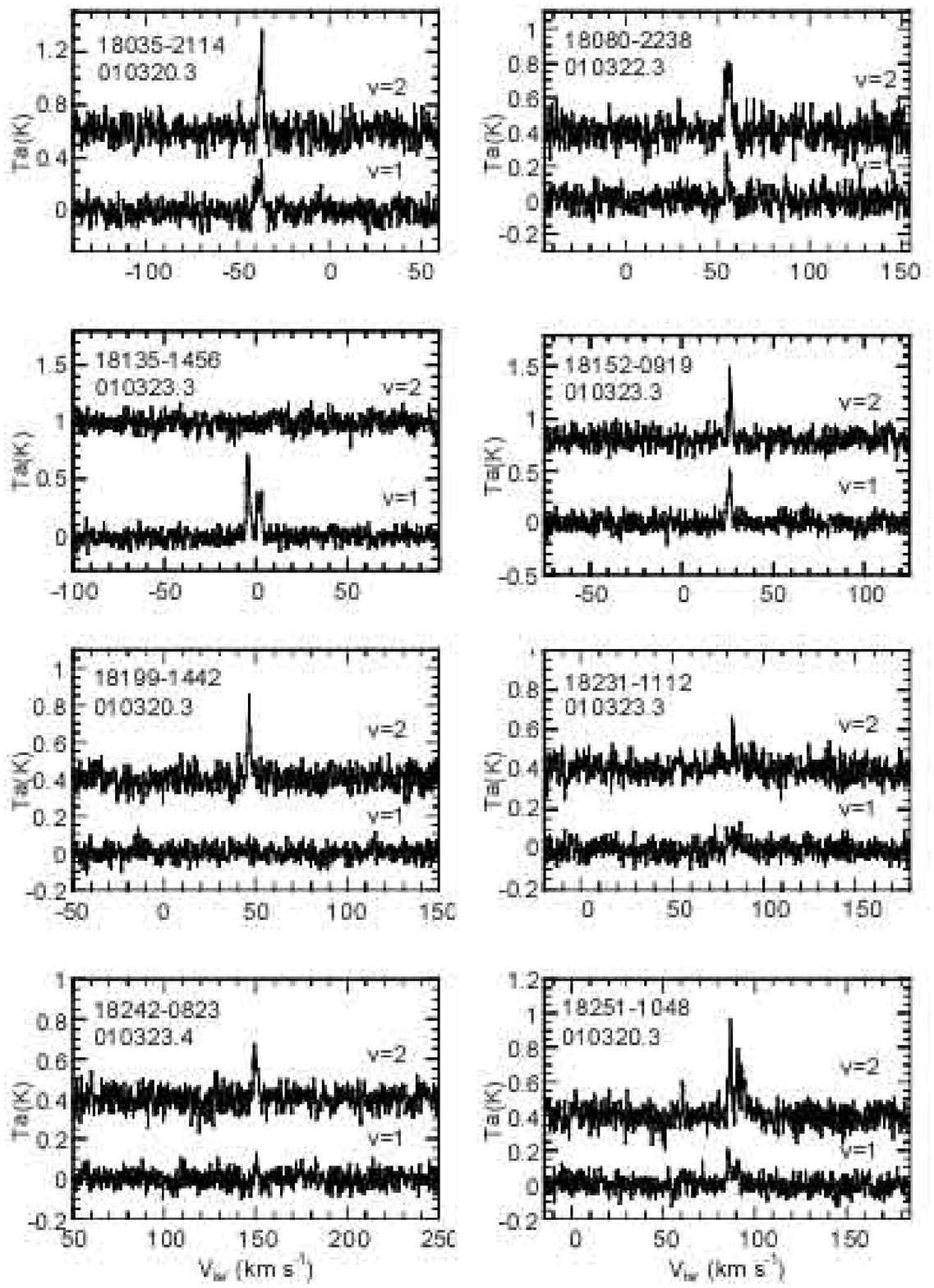}
    %%% \FigureFile(width,height){filename}
  \end{center}
  \caption{Continued.}
  \label{fig:sample}
\end{figure}

\newpage
\renewcommand{\thefigure}{4c}
\begin{figure}
  \begin{center}
    \FigureFile(120mm,10mm){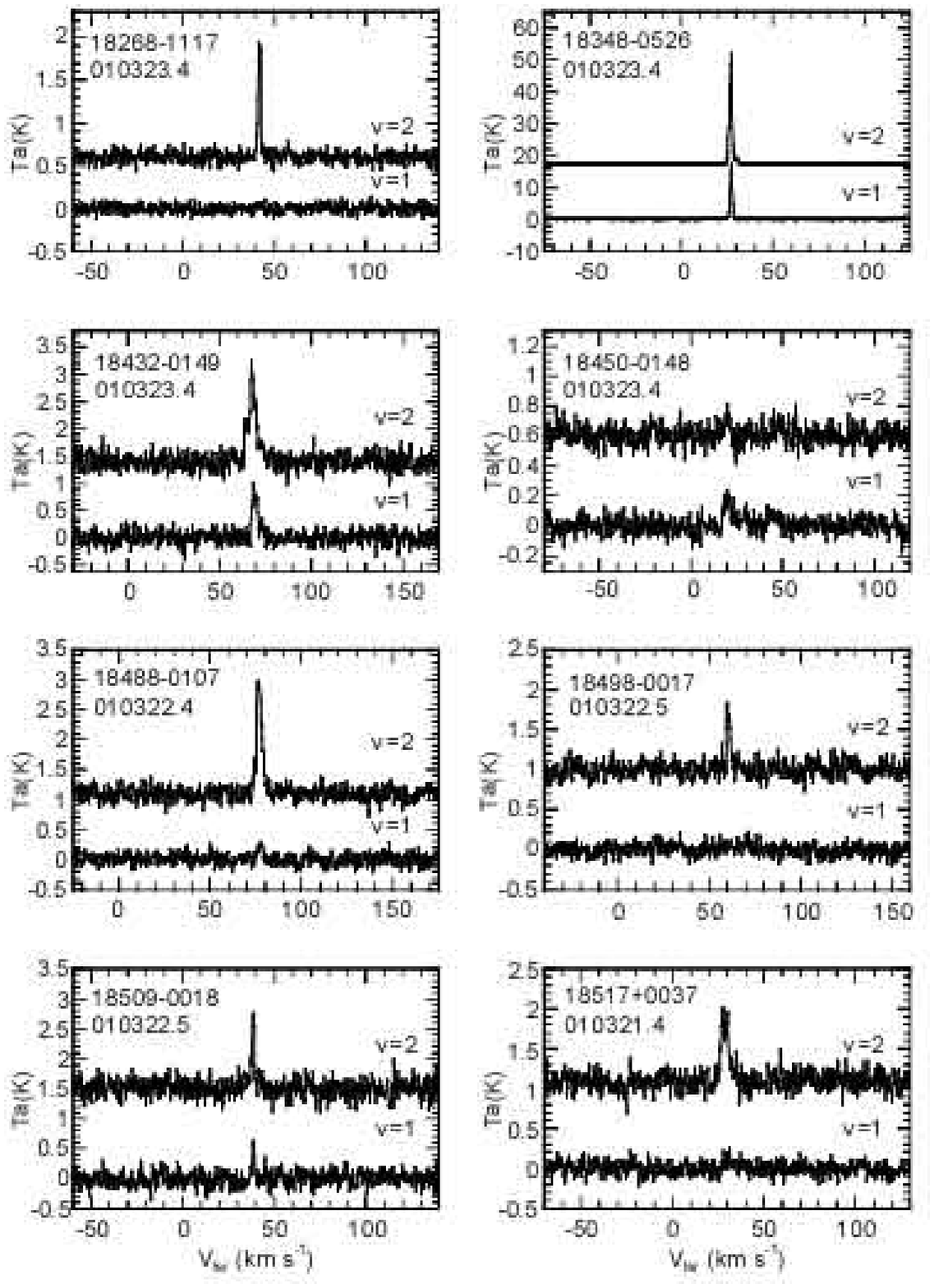}
    %%% \FigureFile(width,height){filename}
  \end{center}
  \caption{Continued.}
  \label{fig:sample}
\end{figure}

\newpage
\renewcommand{\thefigure}{4d}
\begin{figure}
  \begin{center}
    \FigureFile(120mm,10mm){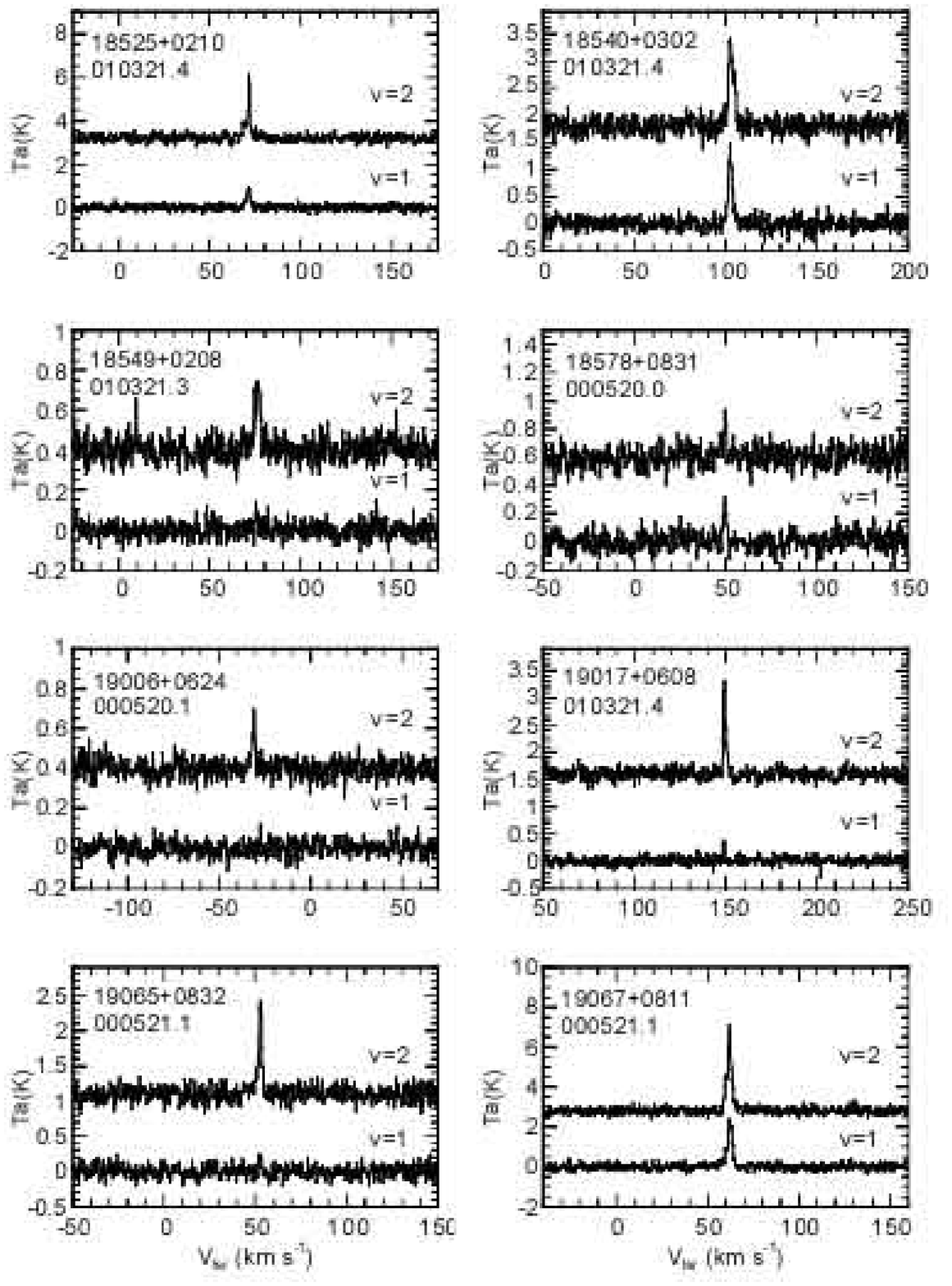}
    %%% \FigureFile(width,height){filename}
  \end{center}
  \caption{Continued.}
  \label{fig:sample}
\end{figure}

\newpage
\renewcommand{\thefigure}{4e}
\begin{figure}
  \begin{center}
    \FigureFile(120mm,10mm){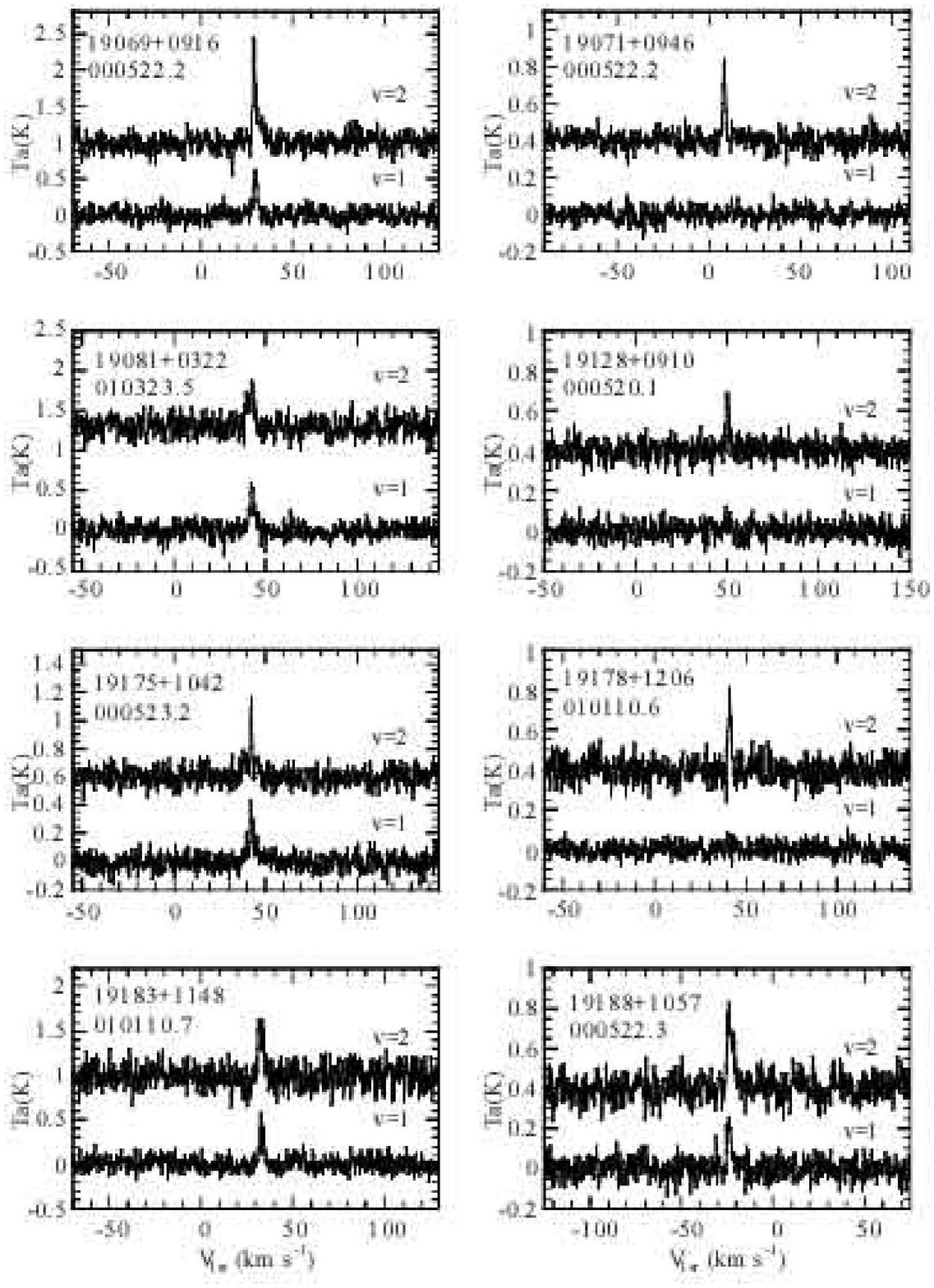}
    %%% \FigureFile(width,height){filename}
  \end{center}
  \caption{Continued.}
  \label{fig:sample}
\end{figure}

\newpage
\renewcommand{\thefigure}{4f}
\begin{figure}
  \begin{center}
    \FigureFile(120mm,10mm){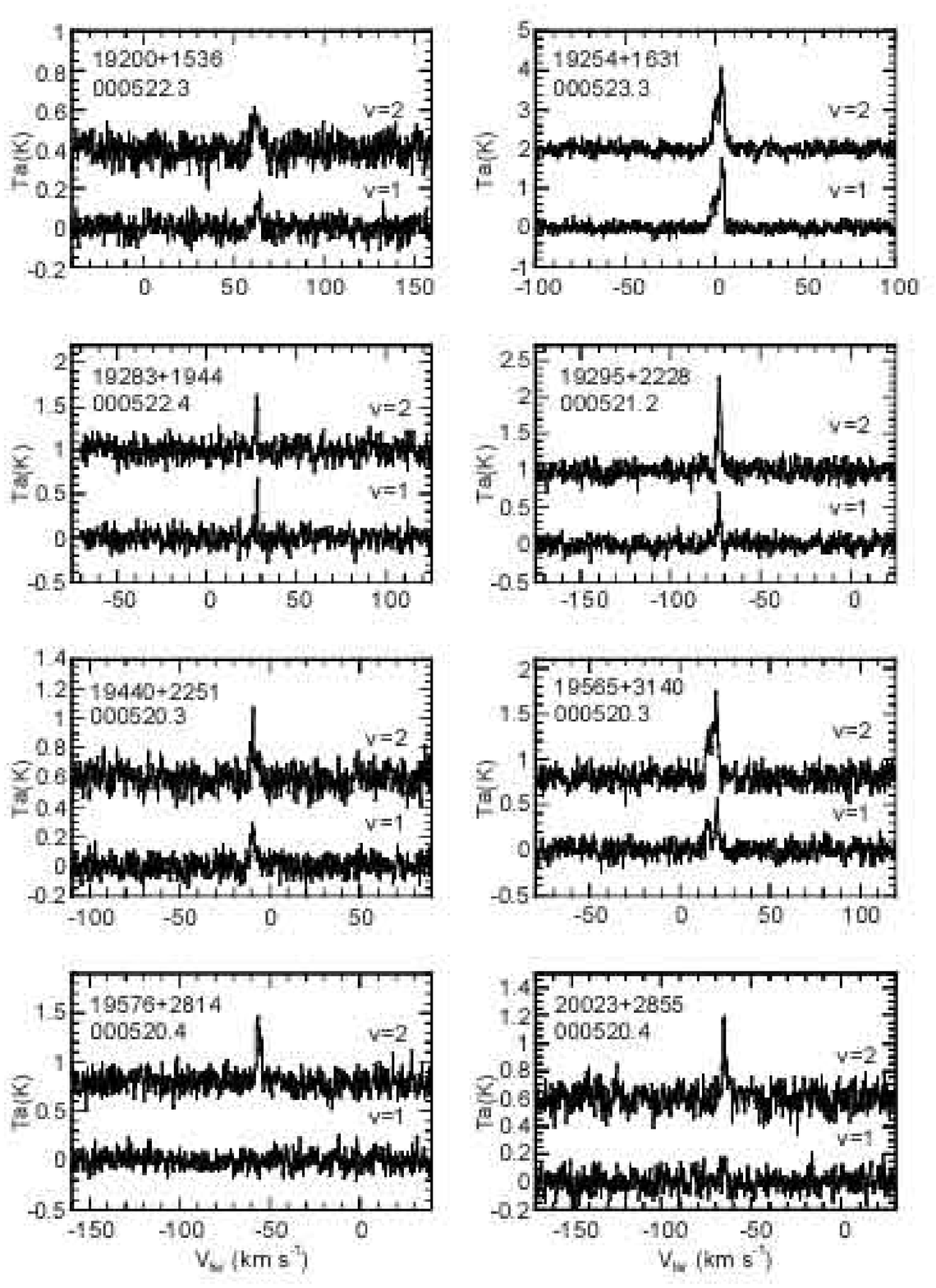}
    %%% \FigureFile(width,height){filename}
  \end{center}
  \caption{Continued.}
  \label{fig:sample}
\end{figure}

\newpage
\renewcommand{\thefigure}{4g}
\begin{figure}
  \begin{center}
    \FigureFile(120mm,10mm){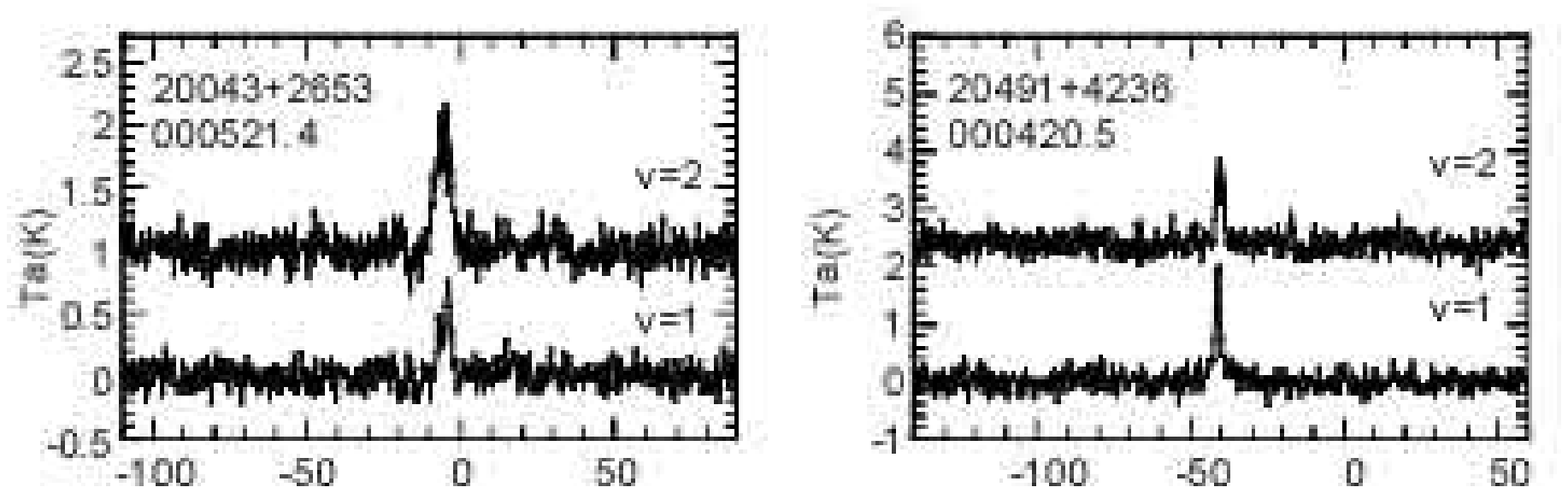}
    %%% \FigureFile(width,height){filename}
  \end{center}
  \caption{Continued.}
  \label{fig:sample}
\end{figure}

\newpage
\renewcommand{\thefigure}{5}
\begin{figure}
  \begin{center}
    \FigureFile(120mm,10mm){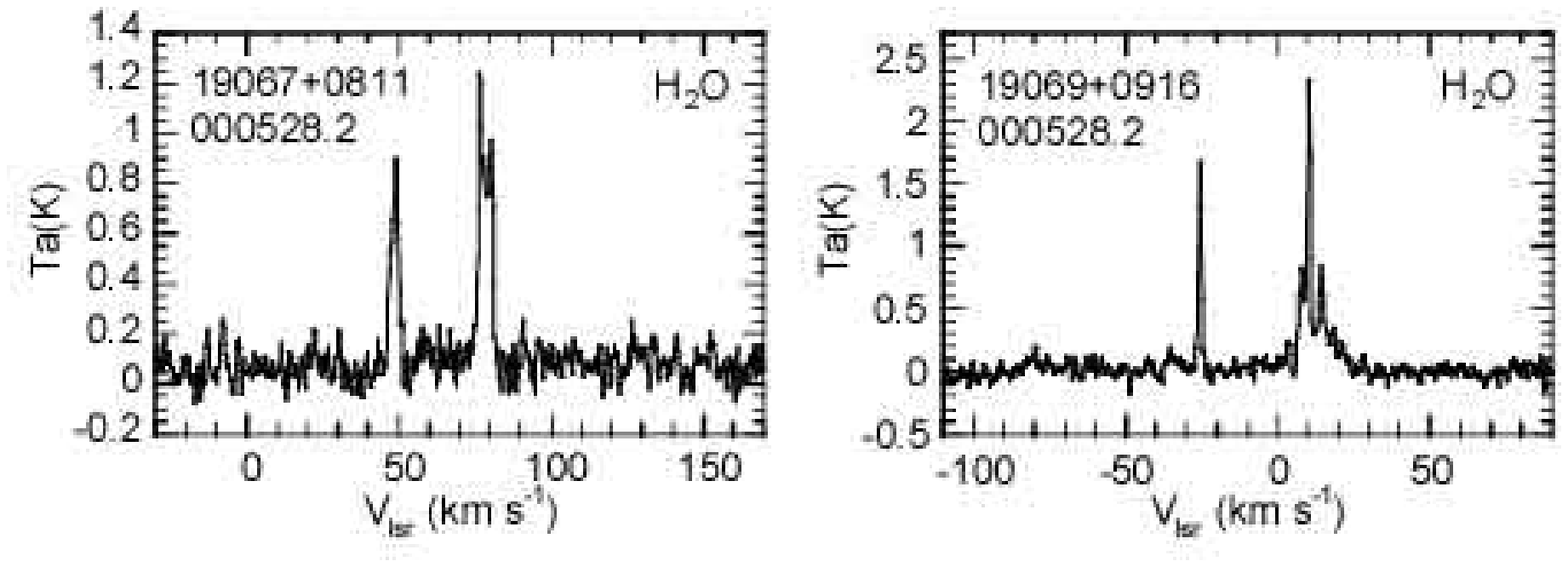}
    %%% \FigureFile(width,height){filename}
  \end{center}
  \caption{Spectra of the H$_2$O maser line at 22 GHz for 2 of 3 
  detected sources except IRAS 19312+1950 (e.g., \cite{nakdeg00}).}\label{fig:sample}
\end{figure}

\newpage
\renewcommand{\thefigure}{6}
\begin{figure}
  \begin{center}
    \FigureFile(120mm,10mm){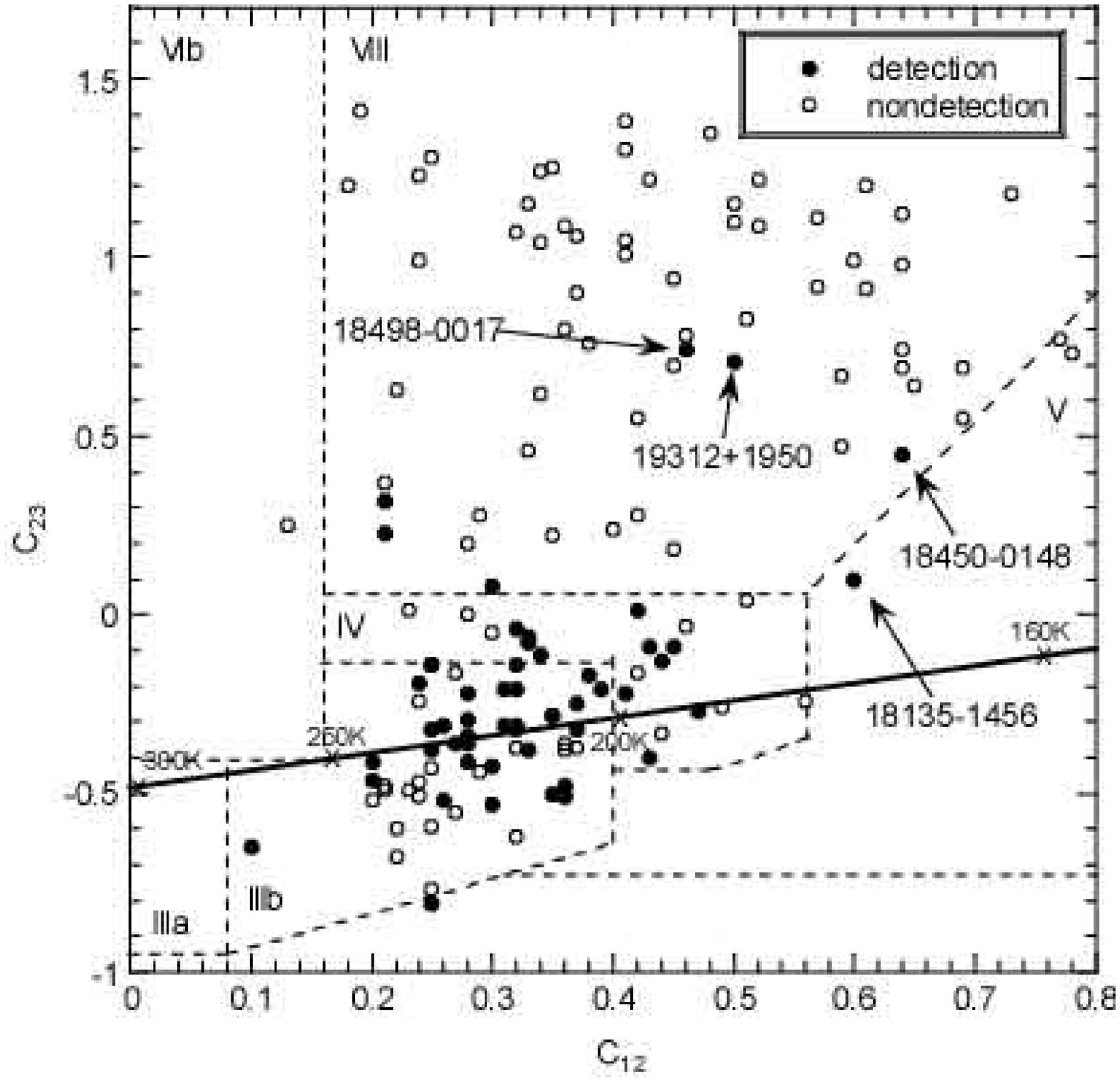}
    %%% \FigureFile(width,height){filename}
  \end{center}
  \caption{IRAS two-color diagram of the present sample. The filled and open circles indicate SiO detections and non-detections, respectively. A black-body curve is indicated by a solid line. The fegions in the diagram indicate a classification of IRAS sources by \citet{van88}. Candidates for post-AGB stars or late-AGB stars with extreme thick envelope are indicated by arrows.}\label{fig:sample}
\end{figure}

\newpage
\renewcommand{\thefigure}{7}
\begin{figure}
  \begin{center}
    \FigureFile(120mm,10mm){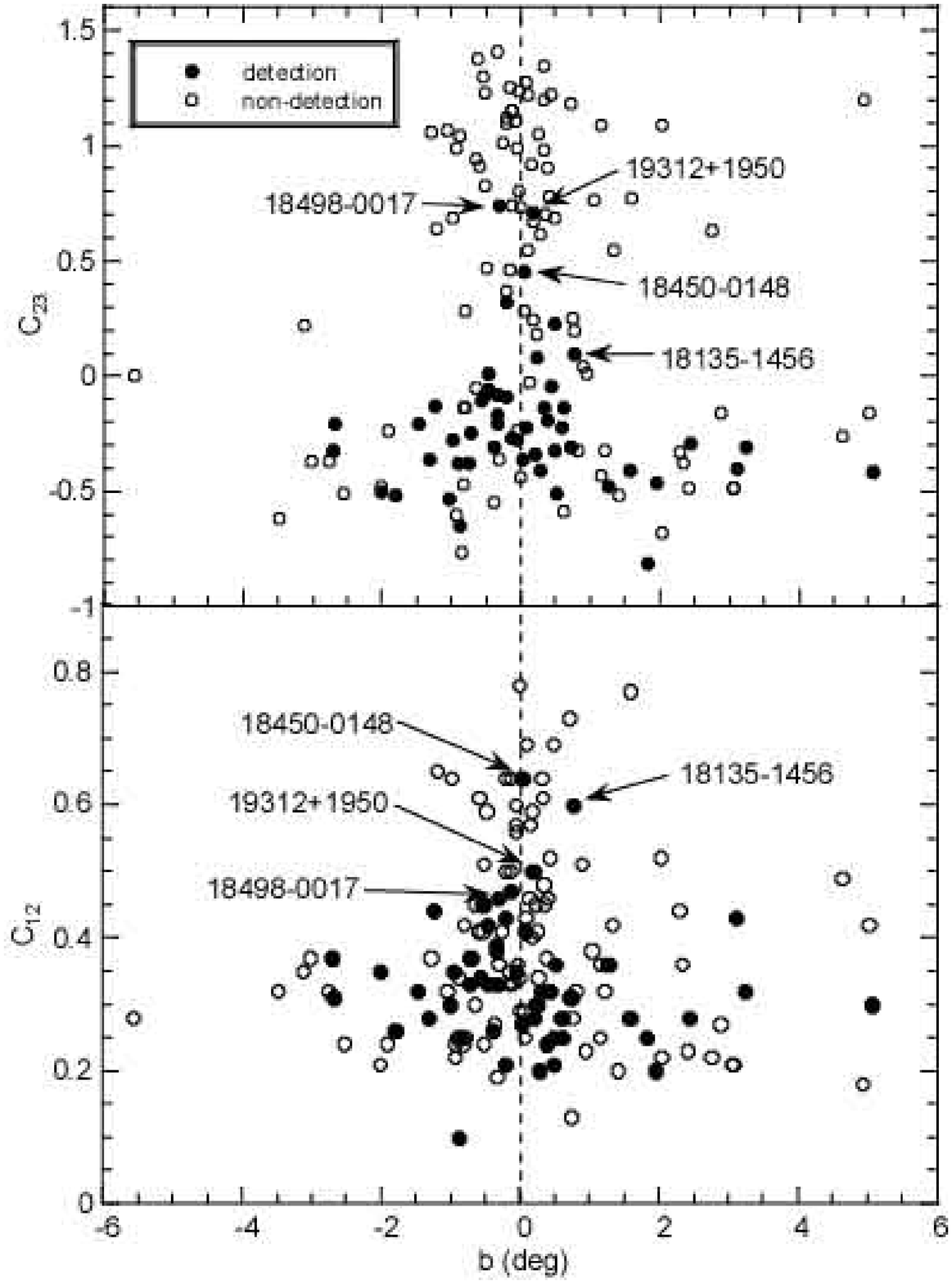}
    %%% \FigureFile(width,height){filename}
  \end{center}
  \caption{Galactic latitude vs. IRAS colors. The filled and open circles are 
  detections and non-detections in SiO, respectively. The names of the candidates   for post-AGB stars (or late-AGB stars) with an extreme-thick envelope are indicated in the diagram.
}\label{fig:sample}
\end{figure}

\newpage
\renewcommand{\thefigure}{8}
\begin{figure}
  \begin{center}
    \FigureFile(120mm,10mm){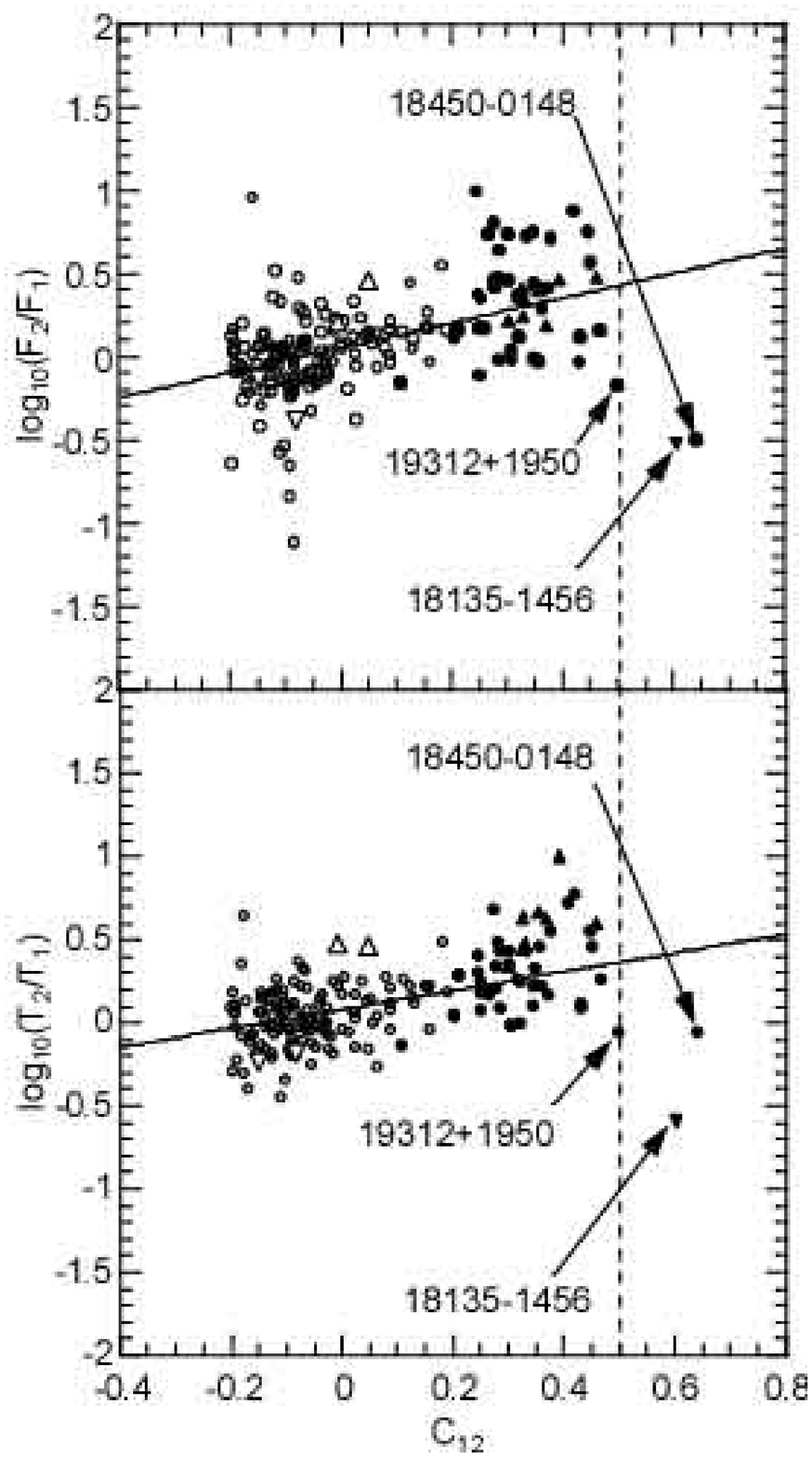}
    %%% \FigureFile(width,height){filename}
  \end{center}
  \caption{IRAS color vs. intensity ratio of the $J=1$--$0$, $v=2$  to the $J=1$--$0$, $v=1$ lines.  In the upper panel, the integrated intensities are used to calculate the intensity ratios. In the lower panel, the peak intensities are used. The filled and open marks indicate the present sample and the sample in previous SiO observations of normal AGB stars, respectively (\cite{nakdeg02}). The triangles and inverse triangles represent the lower and upper limits, respectively. Candidates for post-AGB stars are indicated by arrows.}\label{fig:sample}
\end{figure}

\newpage
\renewcommand{\thefigure}{9}
\begin{figure}
  \begin{center}
    \FigureFile(120mm,10mm){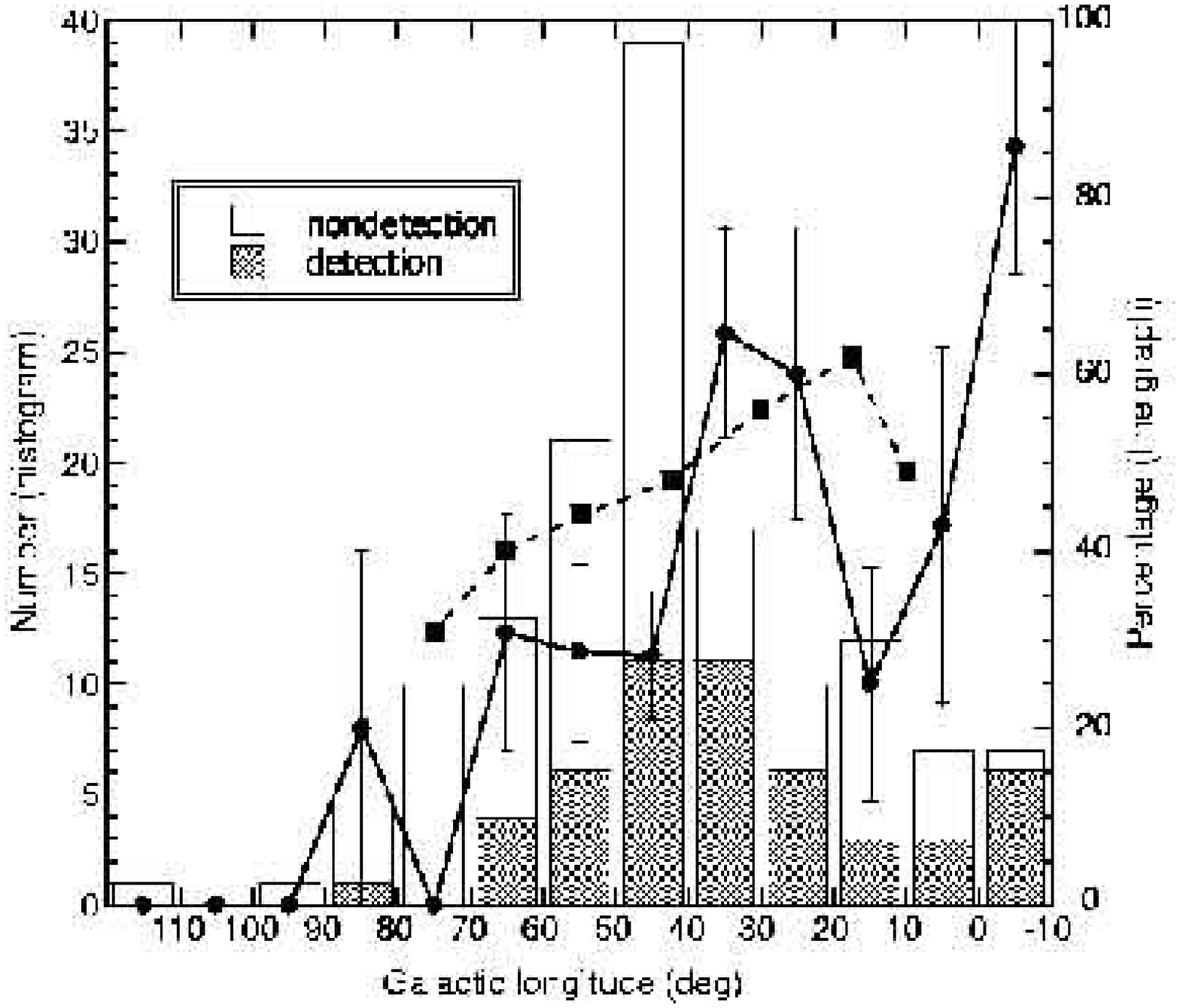}
    %%% \FigureFile(width,height){filename}
  \end{center}
  \caption{Histogram of the present sample with the galactic longitude. The shaded and blank portions in the histogram indicate the number of SiO detections and non-detections, respectively. The line graph drawn by the solid line indicates the detection rate of the present SiO observations.  The line graph drawn by the dotted line indicate the detection rate of an SiO observation for blue IRAS sources with the typical color of normal AGB stars (\cite{nakdeg02}; \cite{nak02}; \cite{izu99}; Deguchi et al. 2000a, b). The error bars indicate the 95\% confidence intervals. }\label{fig:sample}
\end{figure}

\newpage
\renewcommand{\thefigure}{10}
\begin{figure}
  \begin{center}
    \FigureFile(120mm,10mm){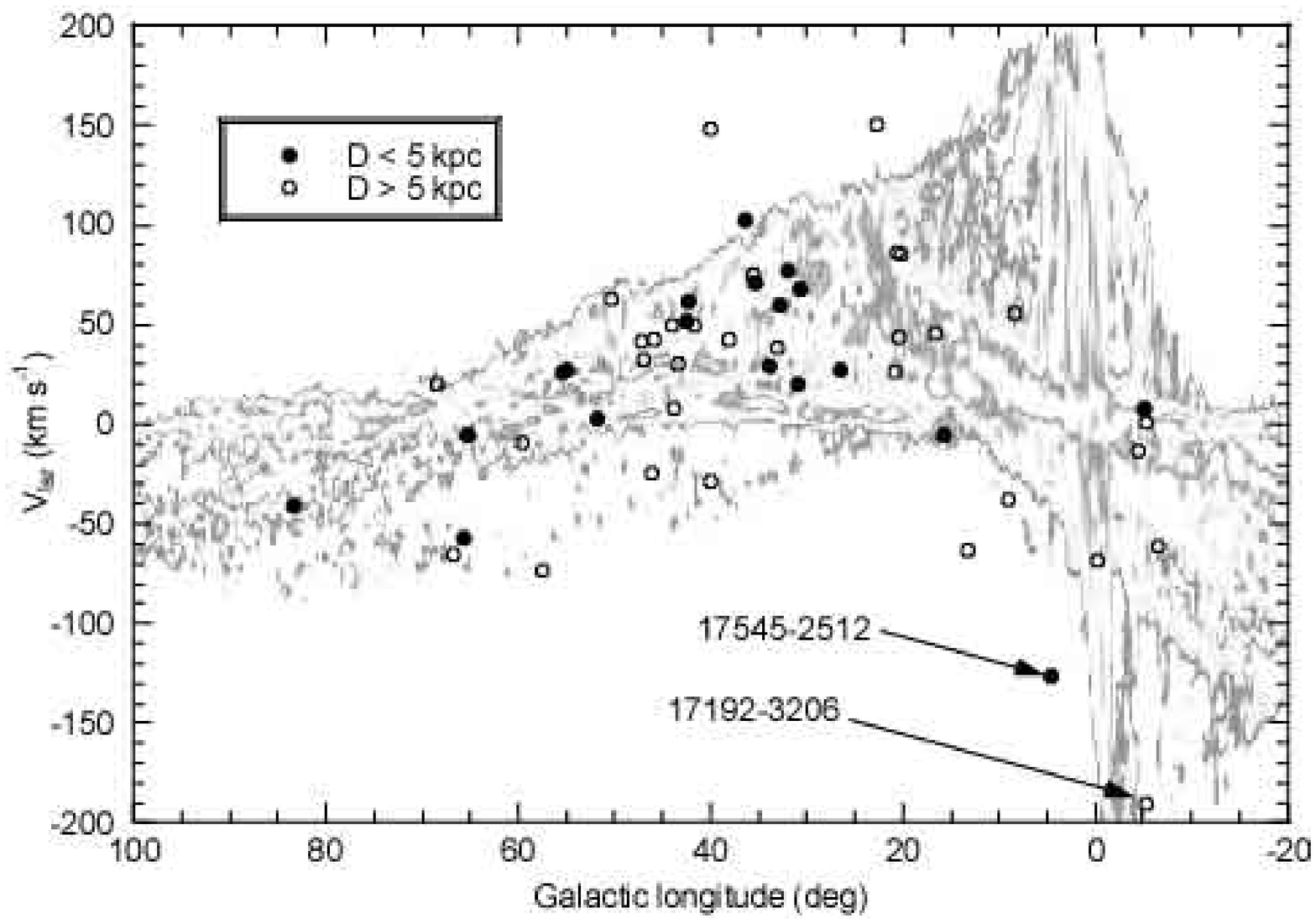}
    %%% \FigureFile(width,height){filename}
  \end{center}
  \caption{Longitude-velocity diagram of the SiO detections overlaid 
  on the CO $v$--$l$ map (taken from \cite{dam01}). The filled and open circles indicate sources at the distance above and below $D=5$ kpc, respectively. The distances used in here are luminosity distance, which is explained in the text. The sources with remarkable velocity are indicated by arrows.}\label{fig:sample}
\end{figure}

\newpage
\renewcommand{\thefigure}{11}
\begin{figure}
  \begin{center}
    \FigureFile(120mm,10mm){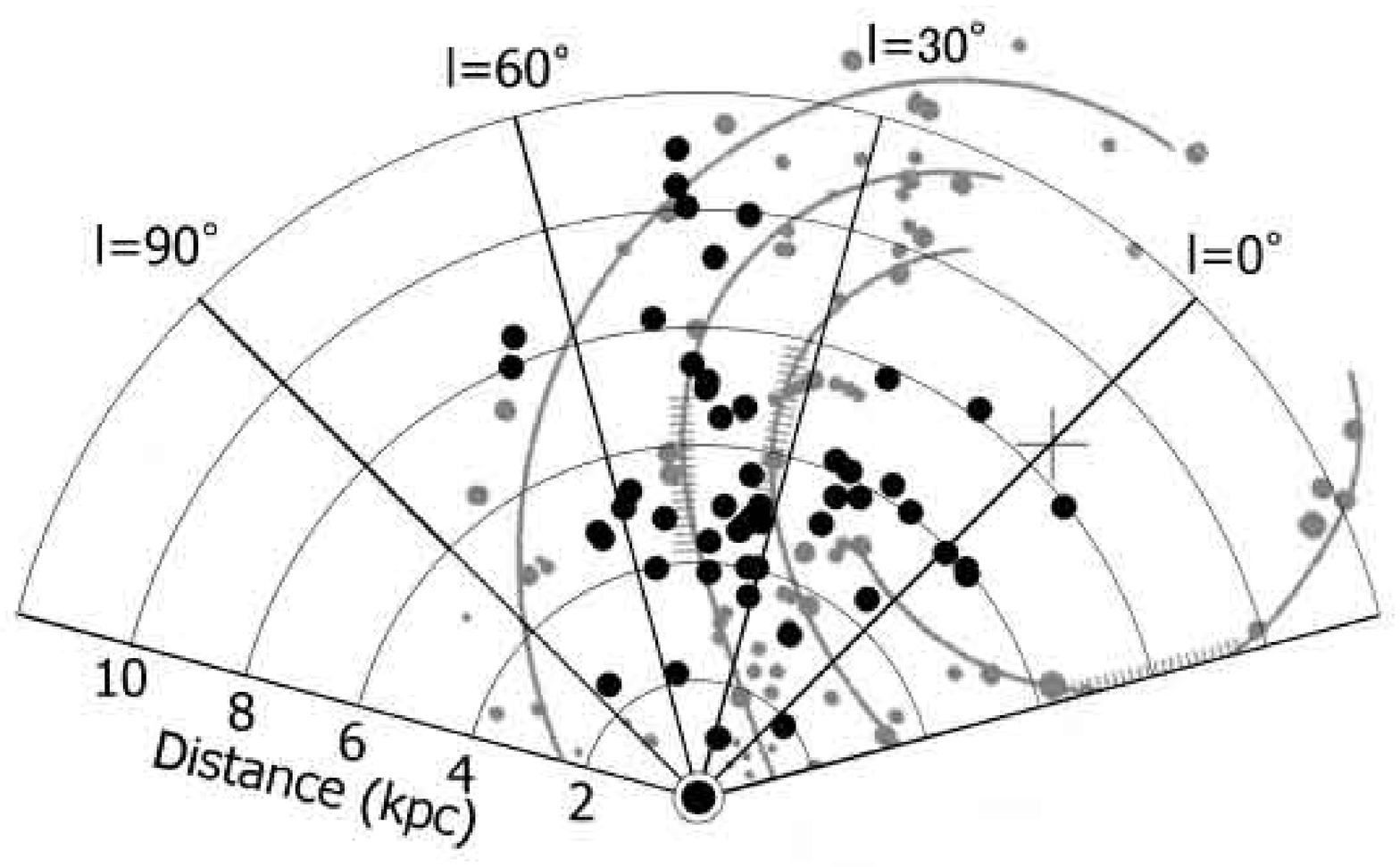}
    %%% \FigureFile(width,height){filename}
  \end{center}
  \caption{Distribution of SiO detected sources projected onto the Galactic plane, overlaid on the spiral arm model of the Galaxy. The filled circles represent SiO sources. The large and small gray circles indicate H{\sc ii} regions, and the hatched areas correspond to the directions of the intensity maxima in the thermal radio continuum and neutral hydrogen emission (\cite{tay93}).}\label{fig:sample}
\end{figure}

%%%%%%%%%%%%%%%%%%
%%%%% TABLES %%%%%
%%%%%%%%%%%%%%%%%%

\include{table1}
\include{table2}
\include{table3}
\include{table4}
\include{table5}
\include{table6}

%\appendix
%\section{Method of .....}

%\section{Approximation of ...}

%\section*{Complete data}

\end{document}

%% file: table1.tex
% TABLE3.tex

{\setlength{\tabcolsep}{2pt}\scriptsize
\begin{longtable}{lrrrrrrrcrc}
\caption{IRAS Data.}
\hline\hline
IRAS name & $l$ & $b$ & $F_{12}$ & $C_{12}$ & $C_{23}$ & $D_{L}$ & $V_{\textrm{lsr}}$ (SiO)  & MSX name & Sep. & 2MASS$^{*}$ \\
 & ($^\circ$) & ($^\circ$) & (Jy) &  & & (kpc) & (km s$^{-1}$) & & ($^{''}$) & \\
\hline
\endhead
\hline
\endfoot
\hline
%\multicolumn{11}{l}{ps: point source, ps$^*$:very bright point source with blue color, \\
%sf: possible star forming region, no: no clear counterpart, \\
%ppn: proto planetary nebula}\\
\multicolumn{11}{l}{*}\\
\multicolumn{11}{l}{ps: point sources, sf: star forming region with extended HII regions,}\\
\multicolumn{11}{l}{sf?: possible star forming region, no: no clear counterpart,}\\
\multicolumn{11}{l}{ppn?: extended sources with PPN-like morphology}\\
\endlastfoot
17192$-$3206 & 354.60 & 2.44 & 11.580  & 0.28 & $-$0.29 & 5.9 & $-$190.3 & 354.5978+02.4362 & 3 & ps \\
17271$-$3309 & 354.66 & 0.48 & 43.800  & 0.69 & 0.69 & 2.5 & $\cdots$ & 354.6629+00.4759 & 6 & sf \\
17286$-$3226 & 355.44 & 0.60 & 11.350  & 0.28 & $-$0.22 & 6.0 & $-$13.2 & 355.4344+00.5970 & 6 & ps \\
17292$-$2727 & 359.66 & 3.23 & 11.550  & 0.32 & $-$0.31 & 5.9 & $-$68.3 & 359.6625+03.2309 & 2 & no \\
17295$-$3321 & 354.76 & $-$0.06 & 11.300  & 0.35 & $-$0.28 & 6.0 & 0.6 & 354.7620$-$00.0612 & 10 & ps \\
17317$-$3331 & 354.88 & $-$0.54 & 104.000  & 0.45 & $-$0.09 & 1.9 & 8.5 & 354.8848$-$00.5381 & 9 & ps \\
17341$-$3529 & 353.50 & $-$2.02 & 6.348  & 0.35 & $-$0.50 & 7.9 & $-$60.6 & 353.5038$-$02.0207 & 18 & ps \\
17528$-$1503 & 13.14 & 5.06 & 10.310  & 0.30 & $-$0.42 & 6.3 & $-$62.7 & 013.1444+05.0638 & 3 & no \\
17545$-$2512 & 4.56 & $-$0.40 & 20.730  & 0.26 & $-$0.31 & 4.4 & $-$126.2 & 004.5628$-$00.3974 & 3 & ps \\
18008$-$2419 & 6.06 & $-$1.20 & 17.630  & 0.65 & 0.64 & 4.1 & $\cdots$ & 006.0569$-$01.1969 & 7 & ps \\
18016$-$2743 & 3.18 & $-$3.02 & 3.022  & 0.37 & $-$0.37 & 11.4 & $\cdots$ & 003.1764$-$03.0243 & 2 & ps \\
18016$-$2540 & 4.97 & $-$2.02 & 13.100  & 0.21 & $-$0.48 & 5.5 & $\cdots$ & 004.9722$-$02.0207 & 3 & no \\
18035$-$2114 & 9.03 & $-$0.21 & 5.957  & 0.21 & 0.32 & 8.2 & $-$37.3 & 009.0299$-$00.2175 & 19 & ps \\
18042$-$2131 & 8.87 & $-$0.49 & 11.230  & 0.59 & 0.47 & 5.3 & $\cdots$ & 008.8729$-$00.4928 & 10 & ps \\
18071$-$1727 & 12.76 & 0.90 & 23.660  & 0.51 & 0.04 & 3.8 & $\cdots$ & 012.7577+00.8951 & 4 & no \\
18080$-$2238 & 8.33 & $-$1.80 & 11.110  & 0.26 & $-$0.52 & 6.1 & 55.5 & 008.3283$$-$$01.8028 & 9 & ps \\
18092$-$1742 & 12.77 & 0.33 & 15.050  & 0.61 & 1.20 & 4.5 & $\cdots$ & 012.7747+00.3351 & 6 & no \\
18112$-$1801 & 12.73 & $-$0.22 & 24.960  & 0.64 & 1.12 & 3.4 & $\cdots$ & 012.7263$-$00.2210 & 18 & sf \\
18125$-$1631 & 14.18 & 0.22 & 10.060  & 0.45 & 0.18 & 6.1 & $\cdots$ & 014.1831+00.2209 & 3 & ps \\
18135$-$1456 & 15.70 & 0.77 & 31.020  & 0.60 & 0.10 & 3.2 & $-$4.9 & 015.7002+00.7711 & 7 & no \\
18152$-$0919 & 20.82 & 3.10 & 14.480  & 0.43 & $-$0.40 & 5.1 & 26.3 & 020.8253+03.0969 & 19 & ps \\
18171$-$1603 & 15.13 & $-$0.53 & 20.080  & 0.51 & 0.83 & 4.2 & $\cdots$ & 015.1296$-$00.5266 & 1 & ps \\
18184$-$1449 & 16.36 & $-$0.21 & 18.420  & 0.50 & 1.10 & 4.4 & $\cdots$ & 016.3609$-$00.2073 & 4 & sf \\
18196$-$1331 & 17.64 & 0.15 & 154.800  & 0.57 & 0.92 & 1.4 & $\cdots$ & 017.6369+00.1568 & 14 & ps \\
18199$-$1442 & 16.65 & $-$0.48 & 12.050  & 0.33 & $-$0.06 & 5.8 & 46.0 & 016.6452$-$00.4833 & 1 & ps \\
18223$-$1243 & 18.66 & $-$0.06 & 16.200  & 0.60 & 0.99 & 4.4 & $\cdots$ & 018.6545$-$00.0582 & 12 & sf \\
18231$-$1112 & 20.09 & 0.49 & 10.590  & 0.21 & 0.23 & 6.1 & 85.0 & 020.0897+00.4860 & 7 & ps \\
18242$-$0823 & 22.71 & 1.58 & 10.660  & 0.28 & $-$0.41 & 6.2 & 150.0 & 022.7076+01.5775 & 10 & ps \\
18251$-$1048 & 20.67 & 0.24 & 6.668  & 0.30 & 0.08 & 7.8 & 86.1 & 020.6720+00.2398 & 3 & ps \\
18262$-$1233 & 19.26 & $-$0.82 & 14.350  & 0.24 & $-$0.47 & 5.3 & $\cdots$ & 019.2580$-$00.8184 & 16 & ps \\
18268$-$1117 & 20.43 & $-$0.34 & 12.370  & 0.39 & $-$0.21 & 5.6 & 43.4 & 020.4323$-$00.3441 & 5 & ps \\
18321$-$0820 & 23.66 & $-$0.13 & 15.960  & 0.64 & 0.74 & 4.3 & $\cdots$ & 023.6582$-$00.1268 & 6 & ps \\
18348$-$0526 & 26.54 & 0.62 & 359.800  & 0.25 & $-$0.14 & 1.1 & 27.2 & 026.5439+00.6178 & 3 & ps \\
18354$-$0638 & 25.55 & $-$0.05 & 18.920  & 0.56 & $-$0.24 & 4.2 & $\cdots$ & 025.5469$-$00.0529 & 2 & ps \\
18418$-$0415 & 28.40 & $-$0.38 & 28.970  & 0.27 & $-$0.55 & 3.8 & $\cdots$ & 028.4037$-$00.3812 & 18 & ps \\
18432$-$0149 & 30.72 & 0.43 & 25.060  & 0.32 & $-$0.04 & 4.0 & 68.2 & 030.7152+00.4268 & 5 & no \\
18434$-$0308 & 29.58 & $-$0.22 & 11.040  & 0.21 & 0.37 & 6.0 & $\cdots$ & 029.5757$-$00.2198 & 12 & ps \\
18450$-$0148 & 30.94 & 0.04 & 23.730  & 0.64 & 0.45 & 3.5 & 20.1 & 030.9444+00.0347 & 8 & ps \\
18488$-$0107 & 31.98 & $-$0.48 & 16.450  & 0.42 & 0.01 & 4.8 & 76.9 & 031.9853$-$00.4854 & 10 & ps \\
18498$-$0017 & 32.83 & $-$0.32 & 22.850  & 0.46 & 0.74 & 4.0 & 60.1 & 032.8286$-$00.3155 & 5 & ps \\
18509$-$0018 & 32.95 & $-$0.57 & 15.650  & 0.34 & $-$0.11 & 5.1 & 38.4 & 032.9535$-$00.5695 & 12 & ps \\
18511+0146 & 34.82 & 0.35 & 23.050  & 0.45 & 0.70 & 4.0 & $\cdots$ & 034.8213+00.3516 & 4 & ps \\
18517+0037 & 33.87 & $-$0.34 & 16.580  & 0.38 & $-$0.17 & 4.9 & 29.6 & 033.8735$-$00.3356 & 4 & no \\
18525+0210 & 35.34 & 0.21 & 18.410  & 0.28 & $-$0.34 & 4.7 & 71.3 & 035.3461+00.2122 & 6 & ps \\
18529+0210 & 35.39 & 0.14 & 5.692  & 0.46 & $-$0.03 & 8.0 & $\cdots$ & 035.3877+00.1349 & 1 & no \\
18540+0302 & 36.29 & 0.28 & 18.570  & 0.20 & $-$0.41 & 4.6 & 102.4 & 036.2929+00.2709 & 28 & ps \\
18542+0114 & 34.71 & $-$0.59 & 10.680  & 0.61 & 0.91 & 5.4 & $\cdots$ & 034.7132$-$00.5952 & 10 & ppn? \\
18549+0208 & 35.58 & $-$0.33 & 13.070  & 0.33 & $-$0.08 & 5.6 & 75.2 & 035.5811$-$00.3335 & 17 & no \\
18578+0831 & 41.59 & 1.96 & 9.353  & 0.20 & $-$0.46 & 6.5 & 49.5 & $\cdots$ & $\cdots$ & ps \\
18587+0521 & 38.87 & 0.32 & 5.432  & 0.64 & 0.98 & 7.4 & $\cdots$ & 038.8744+00.3090 & 24 & no \\
19006+0624 & 40.02 & 0.38 & 4.082  & 0.24 & $-$0.19 & 9.9 & $-$28.9 & $\cdots$ & $\cdots$ & no \\
19008+0530 & 39.25 & $-$0.07 & 12.740  & 0.57 & 1.11 & 5.0 & $\cdots$ & 039.2483$-$00.0649 & 28 & no \\
19011+0852 & 42.27 & 1.41 & 6.284  & 0.20 & $-$0.52 & 7.9 & $\cdots$ & $\cdots$ & $\cdots$ & ps \\
19011+1049 & 44.00 & 2.30 & 9.395  & 0.44 & $-$0.33 & 6.3 & $\cdots$ & $\cdots$ & $\cdots$ & no \\
19017+0608 & 39.91 & 0.02 & 16.500  & 0.27 & $-$0.36 & 5.0 & 148.7 & 039.9158+00.0180 & 8 & ps \\
19025+0613 & 40.07 & $-$0.12 & 3.044  & 0.33 & 1.15 & 11.5 & $\cdots$ & $\cdots$ & $\cdots$ & ps \\
19026+0614 & 40.12 & $-$0.15 & 3.044  & 0.33 & 0.46 & 11.5 & $\cdots$ & $\cdots$ & $\cdots$ & ps \\
19045+0518 & 39.49 & $-$0.99 & 7.395  & 0.64 & 0.69 & 6.3 & $\cdots$ & 039.4939$-$00.9927 & 5 & sf \\
19048+0914 & 43.02 & 0.76 & 3.064  & 0.28 & 0.20 & 11.5 & $\cdots$ & $\cdots$ & $\cdots$ & ps \\
19057+1002 & 43.84 & 0.94 & 3.973  & 0.23 & 0.01 & 10.1 & $\cdots$ & $\cdots$ & $\cdots$ & no \\
19065+0832 & 42.60 & 0.07 & 20.220  & 0.41 & $-$0.22 & 4.4 & 52.2 & $\cdots$ & $\cdots$ & ps \\
19067+0811 & 42.31 & $-$0.13 & 24.630  & 0.47 & $-$0.27 & 3.8 & 62.0 & $\cdots$ & $\cdots$ & ps \\
19069+0916 & 43.28 & 0.32 & 4.798  & 0.32 & $-$0.14 & 9.2 & 30.1 & $\cdots$ & $\cdots$ & ps \\
19069+1335 & 47.12 & 2.33 & 6.851  & 0.36 & $-$0.38 & 7.6 & $\cdots$ & $\cdots$ & $\cdots$ & ps \\
19071+0946 & 43.76 & 0.51 & 7.943  & 0.36 & $-$0.51 & 7.1 & 8.3 & $\cdots$ & $\cdots$ & ps \\
19077+0839 & 42.83 & $-$0.14 & 9.125  & 0.50 & 1.15 & 6.2 & $\cdots$ & $\cdots$ & $\cdots$ & sf \\
19080+0748 & 42.12 & $-$0.61 & 14.140  & 0.41 & 1.38 & 5.2 & $\cdots$ & $\cdots$ & $\cdots$ & sf \\
19081+0322 & 38.19 & $-$2.68 & 9.089  & 0.31 & $-$0.21 & 6.7 & 42.8 & 038.1932$-$02.6822 & 3 & ps \\
19085+0755 & 42.27 & $-$0.65 & 4.630  & 0.30 & $-$0.05 & 9.4 & $\cdots$ & $\cdots$ & $\cdots$ & ps \\
19089+1542 & 49.21 & 2.89 & 23.440  & 0.27 & $-$0.16 & 4.2 & $\cdots$ & $\cdots$ & $\cdots$ & ps \\
19108+0901 & 43.52 & $-$0.65 & 3.689  & 0.45 & 0.94 & 10.0 & $\cdots$ & $\cdots$ & $\cdots$ & ps \\
19112+1220 & 46.50 & 0.81 & 3.697  & 0.32 & $-$0.32 & 10.5 & $\cdots$ & $\cdots$ & $\cdots$ & no \\
19123+1139 & 46.01 & 0.27 & 9.285  & 0.34 & 0.62 & 6.6 & $\cdots$ & $\cdots$ & $\cdots$ & no \\
19128+0910 & 43.88 & $-$1.02 & 8.410  & 0.30 & $-$0.53 & 7.0 & 49.7 & $\cdots$ & $\cdots$ & ps \\
19134+2131 & 54.87 & 4.64 & 5.058  & 0.49 & $-$0.26 & 8.4 & $\cdots$ & $\cdots$ & $\cdots$ & ps \\
19143+1427 & 48.72 & 1.15 & 3.031  & 0.36 & 1.09 & 11.5 & $\cdots$ & $\cdots$ & $\cdots$ & no \\
19167+1313 & 47.90 & 0.06 & 4.183  & 0.29 & 0.28 & 9.9 & $\cdots$ & $\cdots$ & $\cdots$ & no \\
19171+1119 & 46.28 & $-$0.94 & 14.190  & 0.22 & $-$0.60 & 5.3 & $\cdots$ & $\cdots$ & $\cdots$ & ps \\
19175+1042 & 45.78 & $-$1.32 & 7.519  & 0.28 & $-$0.36 & 7.4 & 42.3 & $\cdots$ & $\cdots$ & ps \\
19177+1333 & 48.32 & $-$0.01 & 4.814  & 0.29 & $-$0.44 & 9.2 & $\cdots$ & $\cdots$ & $\cdots$ & ps \\
19178+1206 & 47.06 & $-$0.72 & 3.659  & 0.37 & $-$0.25 & 10.4 & 41.3 & $\cdots$ & $\cdots$ & ps \\
19183+1148 & 46.84 & $-$0.97 & 3.290  & 0.35 & $-$0.28 & 11.0 & 32.7 & $\cdots$ & $\cdots$ & ps \\
19188+1057 & 46.14 & $-$1.47 & 4.020  & 0.32 & $-$0.21 & 10.0 & -24.0 & $\cdots$ & $\cdots$ & ps \\
19195+1650 & 51.42 & 1.16 & 14.610  & 0.25 & $-$0.43 & 5.3 & $\cdots$ & $\cdots$ & $\cdots$ & no \\
19199+2100 & 55.13 & 3.07 & 19.040  & 0.21 & $-$0.49 & 4.6 & $\cdots$ & $\cdots$ & $\cdots$ & no \\
19200+2101 & 55.14 & 3.05 & 30.900  & 0.21 & $-$0.49 & 3.6 & $\cdots$ & $\cdots$ & $\cdots$ & no \\
19200+1536 & 50.37 & 0.48 & 6.059  & 0.25 & $-$0.32 & 8.2 & 63.0 & $\cdots$ & $\cdots$ & ps \\
19200+1035 & 45.96 & $-$1.91 & 3.409  & 0.24 & $-$0.24 & 10.9 & $\cdots$ & $\cdots$ & $\cdots$ & ps \\
19205+1447 & 49.72 & $-$0.02 & 5.760  & 0.34 & 1.24 & 8.4 & $\cdots$ & $\cdots$ & $\cdots$ & sf? \\
19207+1348 & 48.88 & $-$0.52 & 7.349  & 0.24 & 1.23 & 7.4 & $\cdots$ & $\cdots$ & $\cdots$ & ps \\
19207+1346 & 48.85 & $-$0.54 & 4.172  & 0.41 & 1.30 & 9.6 & $\cdots$ & $\cdots$ & $\cdots$ & sf? \\
19211+1432 & 49.57 & $-$0.27 & 4.414  & 0.41 & 1.01 & 9.3 & $\cdots$ & $\cdots$ & $\cdots$ & sf? \\
19223+1359 & 49.23 & $-$0.79 & 6.401  & 0.25 & $-$0.14 & 8.0 & $\cdots$ & $\cdots$ & $\cdots$ & no \\
19226+1401 & 49.29 & $-$0.82 & 6.401  & 0.25 & $-$0.14 & 8.0 & $\cdots$ & $\cdots$ & $\cdots$ & ps \\
19228+1403 & 49.34 & $-$0.86 & 6.401  & 0.25 & $-$0.77 & 8.0 & $\cdots$ & $\cdots$ & $\cdots$ & ps \\
19244+1115 & 47.06 & $-$2.54 & 1346.000  & 0.24 & $-$0.51 & 0.5 & $\cdots$ & $\cdots$ & $\cdots$ & ps \\
19254+1631 & 51.80 & $-$0.22 & 16.570  & 0.43 & $-$0.09 & 4.8 & 3.1 & $\cdots$ & $\cdots$ & ps \\
19270+1550 & 51.40 & $-$0.89 & 6.048  & 0.34 & 1.04 & 8.2 & $\cdots$ & $\cdots$ & $\cdots$ & ps \\
19281+2308 & 57.91 & 2.42 & 4.612  & 0.23 & $-$0.49 & 9.3 & $\cdots$ & $\cdots$ & $\cdots$ & ps \\
19283+1944 & 54.95 & 0.73 & 88.500  & 0.31 & $-$0.31 & 2.1 & 27.8 & $\cdots$ & $\cdots$ & ps \\
19294+1649 & 52.54 & $-$0.93 & 6.694  & 0.24 & 0.99 & 7.8 & $\cdots$ & $\cdots$ & $\cdots$ & ppn? \\
19295+2228 & 57.49 & 1.82 & 14.250  & 0.25 & $-$0.81 & 5.3 & $-$73.0 & $\cdots$ & $\cdots$ & ps \\
19295+1637 & 52.37 & $-$1.05 & 5.051  & 0.32 & 1.07 & 9.0 & $\cdots$ & $\cdots$ & $\cdots$ & no \\
19310+1745 & 53.54 & $-$0.80 & 12.300  & 0.42 & 0.28 & 5.6 & $\cdots$ & $\cdots$ & $\cdots$ & ps \\
19312+1950 & 55.37 & 0.19 & 22.470  & 0.50 & 0.71 & 4.0 & 26.1 & $\cdots$ & $\cdots$ & ppn? \\
19319+2214 & 57.55 & 1.22 & 3.182  & 0.32 & $-$0.32 & 11.3 & $\cdots$ & $\cdots$ & $\cdots$ & ps \\
19327+3024 & 64.78 & 5.02 & 89.390  & 0.42 & $-$0.16 & 2.1 & $\cdots$ & $\cdots$ & $\cdots$ & ps \\
19332+2028 & 56.15 & 0.08 & 6.116  & 0.25 & 1.28 & 8.1 & $\cdots$ & $\cdots$ & $\cdots$ & ps \\
19340+2016 & 56.08 & $-$0.17 & 3.374  & 0.35 & 1.25 & 10.9 & $\cdots$ & $\cdots$ & $\cdots$ & no \\
19341+2038 & 56.41 & $-$0.03 & 5.569  & 0.36 & 0.80 & 8.5 & $\cdots$ & $\cdots$ & $\cdots$ & ps \\
19344+2457 & 60.20 & 2.04 & 4.946  & 0.22 & $-$0.68 & 9.0 & $\cdots$ & $\cdots$ & $\cdots$ & ps \\
19352+2030 & 56.41 & $-$0.31 & 42.890  & 0.36 & $-$0.36 & 3.0 & $\cdots$ & $\cdots$ & $\cdots$ & ps \\
19365+1023 & 47.75 & $-$5.57 & 3.066  & 0.28 & 0.00 & 11.5 & $\cdots$ & $\cdots$ & $\cdots$ & ps \\
19374+1626 & 53.14 & $-$2.76 & 4.067  & 0.32 & $-$0.37 & 10.0 & $\cdots$ & $\cdots$ & $\cdots$ & no \\
19421+2507 & 61.21 & 0.62 & 8.114  & 0.25 & $-$0.59 & 7.1 & $\cdots$ & $\cdots$ & $\cdots$ & no \\
19440+2251 & 59.48 & $-$0.90 & 15.700  & 0.25 & $-$0.38 & 5.1 & $-$9.4 & $\cdots$ & $\cdots$ & no \\
19473+2638 & 63.11 & 0.39 & 6.026  & 0.37 & 0.90 & 8.1 & $\cdots$ & $\cdots$ & $\cdots$ & sf \\
19479+2620 & 62.92 & 0.09 & 4.574  & 0.43 & 1.22 & 9.1 & $\cdots$ & $\cdots$ & $\cdots$ & no \\
19520+2759 & 64.81 & 0.18 & 46.850  & 0.40 & 0.24 & 2.9 & $\cdots$ & $\cdots$ & $\cdots$ & ps \\
19565+3140 & 68.45 & 1.27 & 6.252  & 0.36 & $-$0.48 & 8.0 & 20.5 & $\cdots$ & $\cdots$ & ps \\
19576+2814 & 65.67 & $-$0.74 & 17.410  & 0.33 & $-$0.38 & 4.8 & $-$56.8 & $\cdots$ & $\cdots$ & ps \\
19598+3324 & 70.29 & 1.60 & 302.300  & 0.77 & 0.77 & 0.9 & $\cdots$ & $\cdots$ & $\cdots$ & sf \\
20023+2855 & 66.79 & $-$1.25 & 5.259  & 0.44 & $-$0.13 & 8.4 & $-$65.7 & $\cdots$ & $\cdots$ & ps \\
20028+2903 & 66.96 & $-$1.28 & 20.260  & 0.37 & 1.06 & 4.4 & $\cdots$ & $\cdots$ & $\cdots$ & sf \\
20043+2653 & 65.32 & $-$2.71 & 17.890  & 0.37 & $-$0.32 & 4.7 & $-$4.7 & $\cdots$ & $\cdots$ & ps \\
20072+2710 & 65.91 & $-$3.12 & 3.056  & 0.35 & 0.22 & 11.4 & $\cdots$ & $\cdots$ & $\cdots$ & ps \\
20137+2838 & 67.93 & $-$3.49 & 4.459  & 0.32 & $-$0.62 & 9.5 & $\cdots$ & $\cdots$ & $\cdots$ & ps \\
20187+4111 & 78.87 & 2.76 & 65.290  & 0.22 & 0.63 & 2.5 & $\cdots$ & $\cdots$ & $\cdots$ & sf \\
20197+3722 & 75.84 & 0.41 & 423.700  & 0.46 & 0.78 & 0.9 & $\cdots$ & $\cdots$ & $\cdots$ & sf? \\
20237+4003 & 78.48 & 1.33 & 2.202  & 0.42 & 0.55 & 13.2 & $\cdots$ & $\cdots$ & $\cdots$ & ps \\
20249+3953 & 78.48 & 1.04 & 18.830  & 0.38 & 0.76 & 4.6 & $\cdots$ & $\cdots$ & $\cdots$ & ppn? \\
20255+3932 & 78.26 & 0.74 & 5.300  & 0.13 & 0.25 & 8.2 & $\cdots$ & $\cdots$ & $\cdots$ & ps \\
20288+3934 & 78.68 & 0.26 & 3.501  & 0.41 & 1.05 & 10.5 & $\cdots$ & $\cdots$ & $\cdots$ & ps \\
20305+4010 & 79.34 & 0.34 & 2.461  & 0.48 & 1.35 & 12.1 & $\cdots$ & $\cdots$ & $\cdots$ & ps \\
20319+3958 & 79.35 & 0.00 & 26.610  & 0.78 & 0.73 & 3.0 & $\cdots$ & $\cdots$ & $\cdots$ & sf \\
20321+4112 & 80.35 & 0.72 & 5.227  & 0.73 & 1.18 & 7.1 & $\cdots$ & $\cdots$ & $\cdots$ & sf? \\
20331+4024 & 79.83 & 0.10 & 7.379  & 0.69 & 0.55 & 6.1 & $\cdots$ & $\cdots$ & $\cdots$ & no \\
20333+4102 & 80.36 & 0.44 & 22.850  & 0.52 & 1.22 & 3.9 & $\cdots$ & $\cdots$ & $\cdots$ & sf? \\
20446+4613 & 85.70 & 2.03 & 3.250  & 0.52 & 1.09 & 10.3 & $\cdots$ & $\cdots$ & $\cdots$ & sf? \\
20491+4236 & 83.42 & $-$0.89 & 54.810  & 0.10 & $-$0.65 & 2.4 & $-$40.8 & $\cdots$ & $\cdots$ & ps \\
21080+4758 & 89.64 & 0.17 & 7.490  & 0.59 & 0.67 & 6.5 & $\cdots$ & $\cdots$ & $\cdots$ & sf \\
21329+5113 & 94.79 & $-$0.33 & 2.477  & 0.19 & 1.41 & 12.5 & $\cdots$ & $\cdots$ & $\cdots$ & no \\
23572+6702 & 117.97 & 4.94 & 4.307  & 0.18 & 1.20 & 9.4 & $\cdots$ & $\cdots$ & $\cdots$ & $\cdots$ \\
\end{longtable}}

%% file: table2.tex
% TABLE1.TEX -- Sample table 1.

{\setlength{\tabcolsep}{3pt}\footnotesize
\begin{longtable}{lrrcrrrcrr}
\caption{List of Detections.}
\hline\hline
 & & $J=1$--$0$ & $v=1$ &  & & $J=1$--$0$ & $v=2$ & &   \\
IRAS name & $V_{\textrm{lsr}}$ & $T_{\textrm{a}}$ & S & r.m.s. & $V_{\textrm{lsr}}$ & $T_{\textrm{a}}$ & S & r.m.s. & Obs. date \\
&\multicolumn{4}{c}{------------------------------------------------} & \multicolumn{4}{c}{------------------------------------------------}&\\ 
  & (km s$^{-1}$) & (K) & (K km s$^{-1}$) & (K) & (km s$^{-1}$) & (K) & (K km s$^{-1}$) & (K) & (yymmdd.d) \\
\hline
\endhead
\hline
\endfoot
\hline
\multicolumn{10}{l}{$^{*}$: Sources observed at MSX position (see text). }\\
\multicolumn{10}{l}{$^{**}$: An oject reported by \citet{nakdeg00}. }\\
\multicolumn{10}{l}{(): The values in parentheses are for dubious detections.}\\
\multicolumn{10}{l}{[]: This value is unexpectedly low because of artificial concave at just}\\
\multicolumn{10}{l}{$\,\,\,\,\,$ right side of the maser.}\\
\endlastfoot
17192$-$3206$^{*}$ & $-$189.6 & 0.147 & 0.227 & 0.046 & $-$191.0 & 0.399 & 1.001 & 0.065 & 010322.2 \\
17286$-$3226$^{*}$ & $-$14.1 & 0.123 & 0.128 & 0.033 & $-$12.3 & 0.199 & 0.384 & 0.044 & 010322.2 \\
17292$-$2727$^{*}$ & $\cdots$ & $\cdots$ & $\cdots$ & 0.058 & $-$67.8 & 0.477 & 1.753 & 0.071 & 010322.3 \\
17295$-$3321$^{*}$ & 0.5 & 0.132 & 0.230 & 0.039 & 0.6 & 0.282 & 0.636 & 0.049 & 010322.3 \\
17317$-$3331$^{*}$ & 9.1 & 2.426 & 5.516 & 0.102 & 7.9 & 7.025 & 20.441 & 0.129 & 010323.2 \\
17341$-$3529$^{*}$ & $-$60.6 & 0.212 & 0.487 & 0.066 & $-$60.7 & 0.607 & 0.464 & 0.083 & 010322.2 \\
17528$-$1503$^{*}$ & $\cdots$ & $\cdots$ & $\cdots$ & 0.041 & $-$62.7 & 0.268 & 0.793 & 0.055 & 010322.3 \\
17545$-$2512$^{*}$ & $-$126.1 & 2.299 & 8.126 & 0.056 & $-$126.4 & 3.618 & 12.294 & 0.073 & 010322.3 \\
18035$-$2114$^{*}$ & $-$37.6 & 0.387 & 0.920 & 0.061 & $-$37.1 & 0.753 & 1.321 & 0.080 & 010320.3 \\
18080$-$2238$^{*}$ & 54.4 & 0.279 & 0.271 & 0.053 & 56.6 & 0.416 & 1.509 & 0.064 & 010322.3 \\
18135$-$1456$^{*}$ & $-$4.6 & 0.735 & 2.395 & 0.052 & $\cdots$ & $\cdots$ & $\cdots$ & 0.063 & 010323.3 \\
18152$-$0919$^{*}$ & 26.3 & 0.520 & 1.453 & 0.061 & 26.4 & 0.694 & 1.370 & 0.075 & 010322.4 \\
18199$-$1442$^{*}$ & $\cdots$ & $\cdots$ & $\cdots$ & 0.036 & 46.1 & 0.461 & 0.731 & 0.048 & 010320.3 \\
18231$-$1112$^{*}$ & 87.4 & 0.132 & 0.541 & 0.037 & 82.7 & 0.256 & 0.816 & 0.045 & 010323.3 \\
18242$-$0823$^{*}$ & 150.4 & 0.124 & 0.174 & 0.036 & 149.6 & 0.271 & 0.469 & 0.049 & 010323.4 \\
18251$-$1048$^{*}$ & 85.4 & 0.207 & 0.783 & 0.042 & 86.8 & 0.565 & 2.311 & 0.054 & 010320.3 \\
18268$-$1117$^{*}$ & $\cdots$ & $\cdots$ & $\cdots$ & 0.046 & 41.9 & 1.386 & 1.590 & 0.058 & 010323.4 \\
18348$-$0526$^{*}$ & 27.2 & 17.172 & 26.853 & 0.109 & 27.2 & 34.386 & 65.085 & 0.140 & 010323.4 \\
18432$-$0149$^{*}$ & 68.5 & 1.089 & 3.491 & 0.105 & 67.9 & 1.924 & 8.283 & 0.131 & 010323.4 \\
18450$-$0148$^{*}$ & 20.1 & 0.235 & 1.127 & 0.049 & 20.0 & 0.208 & 0.360 & 0.059 & 010323.4 \\
18488$-$0107$^{*}$ & 76.8 & 0.320 & 0.868 & 0.081 & 76.9 & 1.929 & 6.604 & 0.101 & 010322.4 \\
18498$-$0017$^{*}$ & $\cdots$ & $\cdots$ & $\cdots$ & 0.069 & 60.1 & 0.824 & 2.449 & 0.094 & 010322.5 \\
18509$-$0018$^{*}$ & 38.6 & 0.738 & 0.572 & 0.111 & 38.3 & 1.258 & 3.255 & 0.140 & 010322.5 \\
18517+0037$^{*}$ & 31.4 & 0.265 & 0.727 & 0.077 & 27.7 & 0.952 & 3.828 & 0.103 & 010321.4 \\
18525+0210$^{*}$ & 71.4 & 0.942 & 1.944 & 0.096 & 71.3 & 2.911 & 5.972 & 0.137 & 010321.4 \\
18540+0302$^{*}$ & 102.3 & 1.533 & 4.220 & 0.098 & 102.5 & 1.732 & 5.538 & 0.136 & 010321.4 \\
18549+0208$^{*}$ & 75.2 & 0.142 & 0.245 & 0.037 & 75.2 & 0.384 & 1.329 & 0.054 & 010321.3 \\
18578+0831 & 49.3 & 0.313 & 0.504 & 0.056 & 49.7 & 0.343 & 0.671 & 0.064 & 000520.0 \\
19006+0624 & ($-$27.2) & (0.117) & (0.045) & 0.035 & $-$30.7 & 0.301 & 0.453 & 0.042 & 000520.1 \\
19017+0608$^{*}$ & 148.3 & 0.348 & 0.370 & 0.058 & 149.0 & 1.708 & 2.369 & 0.082 & 010321.4 \\
19065+0832 & 52.0 & 0.250 & [0.007] & 0.079 & 52.3 & 1.329 & 3.025 & 0.091 & 000521.1 \\
19067+0811 & 62.0 & 2.386 & 8.140 & 0.124 & 61.9 & 4.353 & 11.839 & 0.136 & 000521.1 \\
19069+0916 & 30.0 & 0.608 & 0.895 & 0.082 & 30.1 & 0.611 & 1.185 & 0.099 & 000522.2 \\
19071+0946 & $\cdots$ & $\cdots$ & $\cdots$ & 0.035 & 8.3 & 0.484 & 1.048 & 0.040 & 000522.2 \\
19081+0322$^{*}$ & 42.6 & 0.601 & 1.756 & 0.083 & 43.0 & 0.587 & 1.960 & 0.106 & 010323.5 \\
19128+0910 & (48.9) & (0.131) & (0.124) & 0.040 & 50.4 & 0.300 & 0.694 & 0.045 & 000520.1 \\
19175+1042 & 42.3 & 0.453 & 1.297 & 0.057 & 42.2 & 0.562 & 1.261 & 0.065 & 000523.2 \\
19178+1206 & $\cdots$ & $\cdots$ & $\cdots$ & 0.032 & 41.3 & 0.405 & 0.580 & 0.058 & 010110.6 \\
19183+1148 & 32.2 & 0.582 & 0.966 & 0.071 & 33.2 & 0.745 & 0.953 & 0.117 & 010110.7 \\
19188+1057 & $-$23.9 & 0.263 & 0.662 & 0.048 & $-$24.0 & 0.485 & 1.532 & 0.057 & 000522.3 \\
19200+1536 & 64.2 & 0.179 & 0.422 & 0.041 & 61.8 & 0.214 & 0.633 & 0.051 & 000522.3 \\
19254+1631 & 3.2 & 1.775 & 7.519 & 0.095 & 3.0 & 2.197 & 9.960 & 0.111 & 000523.3 \\
19283+1944 & 28.0 & 0.663 & 2.587 & 0.079 & 27.6 & 0.640 & 2.549 & 0.095 & 000522.4 \\
19295+2228 & $-$73.2 & 0.706 & 1.455 & 0.076 & $-$72.8 & 1.268 & 3.292 & 0.089 & 000521.2 \\
19312+1950$^{**}$ & 25.4 & 0.131 & 0.731 & 0.024 & 26.8 & 0.118 & 0.499 & 0.022 & 000531.2 \\
19440+2251 & $-$9.6 & 0.304 & 0.781 & 0.056 & $-$9.1 & 0.472 & 0.612 & 0.070 & 000520.3 \\
19565+3140 & 20.7 & 0.556 & 1.887 & 0.074 & 20.3 & 0.937 & 3.728 & 0.089 & 000520.3 \\
19576+2814 & $\cdots$ & $\cdots$ & $\cdots$ & 0.076 & $-$56.8 & 0.684 & 1.915 & 0.089 & 000520.4 \\
20023+2855 & ($-$66.2) & (0.191) & (0.285) & 0.064 & $-$65.2 & 0.678 & 1.614 & 0.080 & 000520.4 \\
20043+2653 & $-$3.9 & 0.785 & 2.243 & 0.088 & $-$5.5 & 1.186 & 5.904 & 0.099 & 000521.4 \\
20491+4236 & $-$40.8 & 2.045 & 6.321 & 0.129 & $-$40.8 & 1.502 & 4.504 & 0.157 & 000420.5 \\
\end{longtable}}

%% file: table3.tex
% TABLE2.TEX -- Sample table 2.

{\setlength{\tabcolsep}{3pt}\footnotesize
\begin{longtable}{lccr}
\caption{List of non-detections.}
\hline\hline
IRAS name & r.m.s. ($J=$1--0, $v=$1) & r.m.s. ($J=$1--0, $v=$2) & Obs. date \\
 & ------------------------ & ------------------------ & \\ 
 & (K) & (K) & (yymmdd.d) \\
\hline
\endhead
\hline
\endfoot
\hline
\multicolumn{4}{l}{$^{*}$: Sources observed at MSX positions. }\\
\endlastfoot
17271$-$3309$^{*}$ & 0.078 & 0.065 & 010323.2 \\
18008$-$2419$^{*}$ & 0.072 & 0.059 & 010323.2 \\
18016$-$2743$^{*}$ & 0.062 & 0.050 & 010320.3 \\
18016$-$2540$^{*}$ & 0.070 & 0.056 & 010322.3 \\
18042$-$2131$^{*}$ & 0.059 & 0.044 & 010320.3 \\
18071$-$1727$^{*}$ & 0.068 & 0.053 & 010323.2 \\
18092$-$1742$^{*}$ & 0.068 & 0.055 & 010322.4 \\
18112$-$1801$^{*}$ & 0.067 & 0.053 & 010323.2 \\
18125$-$1631$^{*}$ & 0.064 & 0.053 & 010323.3 \\
18171$-$1603$^{*}$ & 0.060 & 0.048 & 010323.3 \\
18184$-$1449$^{*}$ & 0.059 & 0.046 & 010323.3 \\
18196$-$1331$^{*}$ & 0.060 & 0.046 & 010323.3 \\
18223$-$1243$^{*}$ & 0.058 & 0.046 & 010323.3 \\
18262$-$1233$^{*}$ & 0.065 & 0.051 & 010323.4 \\
18321$-$0820$^{*}$ & 0.063 & 0.046 & 010320.4 \\
18354$-$0638$^{*}$ & 0.076 & 0.056 & 010322.4 \\
18418$-$0415$^{*}$ & 0.076 & 0.060 & 010322.4 \\
18434$-$0308$^{*}$ & 0.080 & 0.063 & 010322.4 \\
18511+0146$^{*}$ & 0.084 & 0.061 & 010321.5 \\
18529+0210$^{*}$ & 0.077 & 0.056 & 010321.4 \\
18542+0114$^{*}$ & 0.062 & 0.046 & 010321.4 \\
18587+0521$^{*}$ & 0.089 & 0.072 & 010323.5 \\
19008+0530$^{*}$ & 0.070 & 0.050 & 010321.3 \\
19011+0852 & 0.039 & 0.032 & 000522.1 \\
19011+1049 & 0.058 & 0.049 & 000520.0 \\
19025+0613 & 0.046 & 0.047 & 000524.0 \\
19026+0614 & 0.046 & 0.034 & 010319.4 \\
19045+0518$^{*}$ & 0.085 & 0.063 & 010321.5 \\
19048+0914 & 0.044 & 0.031 & 010320.2 \\
19057+1002 & 0.051 & 0.036 & 010321.3 \\
19069+1335 & 0.045 & 0.056 & 000522.4 \\
19077+0839 & 0.049 & 0.043 & 000520.0 \\
19080+0748 & 0.058 & 0.050 & 000522.2 \\
19085+0755 & 0.061 & 0.051 & 000522.2 \\
19089+1542 & 0.048 & 0.041 & 000521.1 \\
19108+0901 & 0.059 & 0.044 & 010321.3 \\
19112+1220 & 0.045 & 0.036 & 000521.2 \\
19123+1139 & 0.063 & 0.051 & 000522.2 \\
19134+2131 & 0.045 & 0.037 & 000520.1 \\
19143+1427 & 0.052 & 0.037 & 010321.4 \\
19167+1313 & 0.060 & 0.048 & 000522.3 \\
19171+1119 & 0.063 & 0.050 & 000522.3 \\
19177+1333 & 0.043 & 0.035 & 000521.2 \\
19195+1650 & 0.052 & 0.062 & 000522.3 \\
19199+2100 & 0.074 & 0.063 & 000522.3 \\
19199+2100 & 0.064 & 0.078 & 000522.3 \\
19200+2101 & 0.056 & 0.046 & 000523.2 \\
19200+1035 & 0.062 & 0.037 & 010110.7 \\
19205+1447 & 0.056 & 0.047 & 000523.2 \\
19207+1348 & 0.056 & 0.046 & 000523.2 \\
19207+1346 & 0.057 & 0.050 & 000523.2 \\
19211+1432 & 0.059 & 0.048 & 000523.3 \\
19223+1359 & 0.058 & 0.048 & 000523.3 \\
19226+1401 & 0.042 & 0.033 & 000520.1 \\
19228+1403 & 0.061 & 0.054 & 000523.3 \\
19244+1115 & 0.115 & 0.097 & 000420.4 \\
19270+1550 & 0.066 & 0.053 & 000523.3 \\
19281+2308 & 0.044 & 0.036 & 000521.2 \\
19294+1649 & 0.064 & 0.058 & 000523.3 \\
19295+1637 & 0.068 & 0.055 & 000523.3 \\
19310+1745 & 0.071 & 0.063 & 000523.3 \\
19319+2214 & 0.044 & 0.036 & 000521.3 \\
19327+3024 & 0.060 & 0.052 & 000521.3 \\
19332+2028 & 0.043 & 0.036 & 000520.2 \\
19340+2016 & 0.066 & 0.054 & 000521.3 \\
19341+2038 & 0.066 & 0.053 & 000523.4 \\
19344+2457 & 0.045 & 0.038 & 000520.2 \\
19352+2030 & 0.079 & 0.064 & 000521.3 \\
19365+1023 & 0.053 & 0.039 & 010321.4 \\
19374+1626 & 0.059 & 0.052 & 000521.3 \\
19421+2507 & 0.045 & 0.037 & 000520.3 \\
19473+2638 & 0.068 & 0.062 & 000523.4 \\
19479+2620 & 0.048 & 0.041 & 000520.3 \\
19520+2759 & 0.061 & 0.052 & 000521.3 \\
19598+3324 & 0.084 & 0.080 & 000527.4 \\
20028+2903 & 0.130 & 0.106 & 000520.4 \\
20072+2710 & 0.056 & 0.044 & 000520.3 \\
20137+2838 & 0.095 & 0.080 & 000521.4 \\
20187+4111 & 0.068 & 0.059 & 000420.5 \\
20197+3722 & 0.082 & 0.071 & 000427.5 \\
20237+4003 & 0.076 & 0.116 & 000530.3 \\
20249+3953 & 0.113 & 0.098 & 000426.5 \\
20255+3932 & 0.056 & 0.068 & 000530.4 \\
20288+3934 & 0.054 & 0.064 & 000530.3 \\
20305+4010 & 0.060 & 0.067 & 000530.4 \\
20319+3958 & 0.058 & 0.066 & 000529.4 \\
20321+4112 & 0.079 & 0.105 & 000530.3 \\
20331+4024 & 0.072 & 0.080 & 000529.4 \\
20333+4102 & 0.079 & 0.063 & 000529.4 \\
20446+4613 & 0.077 & 0.085 & 000530.3 \\
21080+4758 & 0.073 & 0.065 & 000527.4 \\
21329+5113 & 0.062 & 0.110 & 000531.3 \\
23572+6702 & 0.051 & 0.057 & 000531.4 \\
\end{longtable}}

%% file: table4.tex
% TABLE4.tex

{\setlength{\tabcolsep}{3pt}\footnotesize
\begin{longtable}{lcrrrrrr}
\caption{H$_2$O detections.}
\hline\hline
IRAS name & $\Delta$$V$ & $V_{\textrm{c}}$ & $V_{\textrm{p}}$ & $T_{\textrm{p}}$ & Flux & r.m.s. & Obs. date \\
 & (km s$^{-1}$) & (km s$^{-1}$) & (K) & (K) & (K km s$^{-1}$)& (K) & (yymmdd.d) \\
\hline
\endhead
\hline
\endfoot
\hline
\endlastfoot
19067+0811 & 27.8  & 62.8 & 76.7 & 1.243 & 8.482 & 0.053 & 000528.2 \\
 &  &  & 48.9 & 0.897 & 7.239 &  &  \\
19069+0916 & 36.4  & $-$7.5 & 10.7 & 2.328 & 35.166 & 0.049 & 000528.2 \\
 &  &  & $-$25.7 & 1.675 & 8.801 &  &  \\
19312+1950 &  & 17.4 & 17.4 & 2.339 & 7.455 & 0.062 & 000601.0 \\
\end{longtable}}

%% file: table5.tex
% TABLE5.tex

{\setlength{\tabcolsep}{3pt}\footnotesize
\begin{longtable}{lcr}
\caption{H$_2$O non-detections.}
\hline\hline
IRAS name & r.m.s. & Obs. date \\
 & (K) & (yymmdd.d) \\
\hline
\endhead
\hline
\endfoot
\hline
\endlastfoot
18171$-$1603 & 0.055 & 000528.1 \\
18578+0831 & 0.080 & 000528.2 \\
19006+0624 & 0.081 & 000528.2 \\
19065+0832 & 0.067 & 000528.2 \\
19071+0946 & 0.077 & 000528.2 \\
19128+0910 & 0.081 & 000528.3 \\
19175+1042 & 0.080 & 000528.3 \\
19178+1206 & 0.023 & 010210.0 \\
19183+1148 & 0.023 & 010210.0 \\
19188+1057 & 0.096 & 000528.3 \\
19200+1536 & 0.101 & 000528.3 \\
19254+1631 & 0.066 & 000531.9 \\
19283+1944 & 0.052 & 000601.0 \\
19295+2228 & 0.073 & 000601.0 \\
19309+2646 & 0.033 & 010210.0 \\
19440+2251 & 0.090 & 000528.3 \\
19565+3140 & 0.088 & 000528.3 \\
19576+2814 & 0.087 & 000528.3 \\
20023+2855 & 0.073 & 000601.0 \\
20043+2653 & 0.145 & 000528.3 \\
\end{longtable}}

%% file: table6.tex
% TABLE6.tex

{\setlength{\tabcolsep}{3pt}\footnotesize
\begin{longtable}{lrclclcl}
\caption{detected masers.}
\hline\hline
IRAS name & \multicolumn{3}{c}{OH (1612 MHz)}  & \multicolumn{2}{c}{H$_2$O}  & \multicolumn{2}{c}{SiO, $J=1$--$0$, $v=1,2$} \\
  & \multicolumn{3}{c}{---------------------------------} 
  & \multicolumn{2}{c}{------------------} & \multicolumn{2}{c}{------------------} \\
 & $V_{\textrm{lsr}}$ & $\Delta$$V$ & Ref. & $V_{\textrm{lsr}}$ & Ref. & $V_{\textrm{lsr}}$ & Ref.  \\
 & {\tiny (km s$^{-1}$)} &{\tiny (km s$^{-1}$)} &  & {\tiny (km s$^{-1}$)} &  & {\tiny (km s$^{-1}$)} & \\
\hline
\endhead
\hline
\endfoot
\hline
\multicolumn{8}{l}{References: 1---this paper, 2---\citet{ces88}, 3---\citet{che93}, 
}\\
\multicolumn{8}{l}{4---\citet{dav93}, 5---\citet{ede88}, 6---\citet{eng96}, 
}\\
\multicolumn{8}{l}{7---\citet{eng86}, 8---\citet{jew91}, 9---\citet{lik89}, 
}\\
\multicolumn{8}{l}{10---\citet{nym93}, 11---\citet{sev97}, 12---\citet{tel89},
}\\
\multicolumn{8}{l}{13--\cite{tel91a}a.}
\endlastfoot
17192$-$3206 & & & &  &  & $-$190.3 & 1 \\
17286$-$3226 & $-$12.5 & 30.5  & 11, 13 &  &  & $-$13.2 & 1 \\
17292$-$2727 & $-$67.2 & 35.2  & 4, 13 &  &  & $-$68.3 & 1 \\
17295$-$3321 & 0.6  & 30.1  & 11, 12, 13 &  &  & 0.6 & 1 \\
17317$-$3331 & 8.6  & 29.2  & 4, 11, 12  &  &  & 8.5 & 1, 10 \\
17341$-$3529 & & & &  &  & $-$60.6 & 1 \\
17528$-$1503 & $-$67.5 & 29.0  & 12 &  &  & $-$62.7 & 1 \\
17545$-$2512 & $-$123.5 & 45.2  & 4, 11, 12 &  &  & $-$126.2 & 1 \\
18016$-$2743 & 72.5  & 22.0  & 4, 11, 13 &  &  &  &  \\
18035$-$2114 & &  & &  &  & $-$37.3 & 1 \\
18042$-$2131 & 193.3  & 33.6  & 11 &  &  &  &  \\
18071$-$1727 & 25.9  & 16.9  & 4, 12 &  &  &  &  \\
18080$-$2238 & 55.0  & 33.6  & 11, 13 &  &  & 55.5 & 1 \\
18092$-$1742 & 138.5  & 27.0  & 12 &  &  &  &  \\
18135$-$1456 & $-$1.15 & 28.2  & 4, 12 & y & 7 & $-$4.9 & 1, 10 \\
18152$-$0919 & 26.7  & 28.2  & 4, 12 &  &  & 26.3 & 1 \\
18196$-$1331 & 31.0  & 18.0  & 12 &  &  &  &  \\
18199$-$1442 & 40.6  & 7.7  & 13 &  &  & 46.0 & 1 \\
18231$-$1112 &  &  & &  &  & 85.0 & 1 \\
18242$-$0823 &  &  &  &  &  & 150.0 & 1 \\
18251$-$1048 & 91.0  & 40.0  & 12 &  &  & 86.1 & 1 \\
18268$-$1117 & 42.5  & 34.5  & 4, 13 &  &  & 43.4 & 1 \\
18348$-$0526 & 27.8  & 27.9  & 4, 12, 13 & 35.0 & 2 & 27.2 & 1, 10 \\
18432$-$0149 & 67.1  & 35.1  & 4, 12, 13 & 55.0 & 2 & 68.2 & 1 \\
18434$-$0308 & 76.2  & 41.0  & 13 &  &  &  &  \\
18450$-$0148 & 34.0  & 14.0  & 12 &  &  & 20.1 & 1 \\
18488$-$0107 & 75.8  & 40.9  & 4, 12 &  &  & 76.9 & 1 \\
18498$-$0017 & 60.7  & 30.8  & 12 & 48.0 & 2, 7 & 60.1 & 1 \\
18509$-$0018 &  &  &  &  &  & 38.4 & 1 \\
18517+0037 & 28.0  & 34.2  & 4, 13 & 28.4 & 6 & 29.6 & 1 \\
18525+0210 & 70.0  & 38.6  & 13, 4 & 70.7 & 6 & 71.3 & 1 \\
18540+0302 & 102.9  & 37.4  & 12, 13 & 85.8 & 2, 6 & 102.4 & 1 \\
18549+0208 & 78.6  & 27.4  & 4, 13 & y & 6 & 75.2 & 1 \\
18578+0831 & 49.3  & 29.0  & 13 & y & 6 & 49.5 & 1 \\
19006+0624 & $-$29.9 & 30.4  & 4, 13 &  &  & $-$28.9 & 1 \\
19017+0608 & 149.6  & 28.5  & 4, 13 & y & 6 & 148.7 & 1, 8 \\
19065+0832 & 53.1  & 35.4  & 4, 12 &  &  & 52.2 & 1 \\
19067+0811 & 59.5  & 32.5  & 4, 12 & y & 6 & 62.0 & 1,10, \\
    & &&&&&& 8\\
19069+0916 & 31.7 & 43.6 & 5  & 26.7 & 2, 6 & 30.1 & 1 \\
19071+0946 & 9.1  & 23.2  & 4, 12 & y & 6, 7 & 8.3 & 1 \\
19081+0322 & 42.2  & 32.0  & 4, 13 & 43.0 & 6 & 42.8 & 1 \\
19085+0755 & 77.3  & 38.9  & 4, 13 & 78.1 & 6 &  &  \\
19112+1220 & 62.2  & 33.1  & 4 &  &  &  &  \\
19128+0910 & 52.4  & 25.3  & 12, 13 &  &  & 49.7 & 1 \\
19134+2131 &  &  &  & $-$17.2 & 2 &  &  \\
19175+1042 &  &  &  & y & 6 & 42.3 & 1 \\
19178+1206 &  &  &  &  &  & 41.3 & 1 \\
19183+1148 & 35.5  & 31.8  & 13 &  &  & 32.7 & 1 \\
19188+1057 & $-$21.2 & 27.2  & 13 &  &  & $-$24.0 & 1 \\
19199+2100 & 56.3  & 26.2  & 13 &  &  &  &  \\
19200+2101 & 55.3  & 26.7  & 12 &  &  &  &  \\
19200+1536 & 61.0  & 26.2  & 4, 12, 13 &  &  & 63.0 & 1 \\
19244+1115 & 76.4  & 61.4  & 12, 13 &  &  &  &  \\
19254+1631 & 1.4  & 37.2  & 4, 12 &  &  & 3.1 & 1 \\
19283+1944 & 27.0  & 28.0  & 12 &  &  & 27.8 & 1 \\
19295+2228 & $-$74.5 & 25.0  & 12 &  &  & $-$73.0 & 1 \\
19312+1950 &  &  &  &  &  & 26.1 & 1 \\
19319+2214 & 20.6  & 16.3  & 4 & 14.2 & 6 &  &  \\
19344+2457 & 16.1  & 22.9  & 13 &  &  &  &  \\
19352+2030 & 0.5  & 7.0  & 12 &  &  &  &  \\
19374+1626 & $-$28.6 & 26.1  & 4, 13 &  &  &  &  \\
19440+2251 & $-$7.58 & 34.9  & 4, 12 &  &  & $-$9.4 & 1 \\
19565+3140 &  &  &  & y & 6 & 20.5 & 1 \\
19576+2814 & $-$58.87 & 21.5  & 4, 9, 12 &  &  & $-$56.8 & 1 \\
19598+3324 &  &  &  & $-$19 & 2 &  &  \\
20023+2855 & $-$63.65 & 25.6  & 4, 12 &  &  & $-$65.7 & 1 \\
20043+2653 & $-$4.56 & 26.4  & 4, 9 &  &  & $-$4.7 & 1 \\
20072+2710 & n &  & 3 &  &  &  &  \\
20137+2838 &  &  &  & $-$67.9 & 2, 6 &  &  \\
20491+4236 & $-$39 & 36.0  & 12 &  &  & $-$40.8 & 1 \\
\end{longtable}}